\documentclass{aa}

\usepackage{natbib,twoopt}
\usepackage{graphicx}
\usepackage{float}
\usepackage{txfonts}
\usepackage{multirow}
\usepackage{tablefootnote}
\usepackage{amsmath}
\usepackage{subcaption}
\usepackage[dvipsnames]{xcolor}
\usepackage{tikz}

\usetikzlibrary{shapes, arrows.meta, positioning, matrix, calc}
\definecolor{darkgoldenrod}{rgb}{0.72, 0.53, 0.04}

\usepackage[colorlinks = true,
            linkcolor = RoyalBlue,
            urlcolor = darkgoldenrod,
            citecolor = RoyalBlue,
            anchorcolor = RoyalBlue, 
            draft = false]{hyperref}

\DeclareRobustCommand{\rev}[1]{{ #1}}
\newenvironment{revisionblock}{\begingroup}{\par\endgroup}
\newcommand{\comment}[1]{}

\definecolor{diagramdarkblue}{rgb}{0,0,0.55}
\definecolor{diagramdarkred}{rgb}{0.55,0,0}

\begin{document}
   \title{Revealing a multi-zone circumnuclear dust response in UGC~11487 through phase-resolved WISE W1--W2 diagnostics}

\author{
I. O. Izviekova\inst{1,2}\corrauth{izviekova@gmail.com}
\and
O. V. Kompaniiets\inst{1}
\and
I. B. Vavilova\inst{1}
\and
D. V. Dobrycheva\inst{1}
}

\institute{
Main Astronomical Observatory of the NAS of Ukraine,
27 Akademik Zabolotny St., Kyiv, 03143, Ukraine
\and
International Centre for Astronomical, Medical and Ecological Research,
27 Akademik Zabolotny St., Kyiv, 03143, Ukraine
}

\titlerunning{Phase-dependent MIR dust response in the TDE candidate UGC~11487}
\authorrunning{Izviekova et al.}

   \date{Received July 2026; accepted to A\&A 1 November 2026}

\abstract
{Mid-infrared (MIR) flares trace dust reprocessing in obscured tidal disruption event (TDE) candidates, especially when the primary optical or ultraviolet flare is weak, missed, or strongly obscured.}
{We investigate the phase-dependent $W1$--$W2$ evolution of the MIR-selected TDE candidate UGC~11487/WTP14adeqka and test whether its infrared behaviour can be described by a single self-similar MIR response or requires a structured circumnuclear dust response.}
{For the first time, we performed a phase-resolved analysis of the WISE/NEOWISE W1--W2 evolution of UGC~11487/WTP14adeqka over 2014--2024. We combined empirical $W1$--$W2$ timing, colour, and flux diagnostics with phenomenological modeling and axisymmetric dust radiative-transfer calculations driven by hydrodynamically motivated fallback histories. We also constructed an aperture-matched ultraviolet-to-radio spectral energy distribution of the host galaxy.}
{UGC~11487 exhibits a decade-long MIR flare characterized by a delayed and broader $W2$ response, colour hysteresis, and systematic cooling from $\sim900$ K to $\sim500$--600 K. Observed-window positive-fluence centroids give a $W2$--$W1$ centroid offset of $\Delta t_{\rm cent}=224\pm45$ d, corresponding to an effective differential response scale of $0.188\pm0.038$ pc. The effective infrared luminosity reaches $\sim3\times10^{43}$ erg s$^{-1}$, and the integrated IR energy is $(2.9$--$3.8)\times10^{51}$ erg. A single fixed-colour phenomenological component is strongly disfavored ($\Delta{\rm BIC}=281.7$), while the preferred compact description is a two-component phenomenological model. Two physical dust-echo solutions with different fallback histories reproduce the broad W1 and W2 light-curve evolution through compact and extended dust populations. The host is a moderately star-forming isolated barred disc, modeled under an absent or weak persistent AGN assumption.}
{The MIR evolution supports an obscured
TDE-like accretion transient reprocessed by a
spatially and thermally distributed dust population.
Phase-resolved WISE diagnostics constrain differential
interband response scales, while radiative-transfer
modeling connects the observed evolution to the
source history and dust geometry.
The two-band data admit multiple physical solutions.}

   \maketitle
   \nolinenumbers

\section{Introduction}
\label{sec:introduction}

Tidal disruption events (TDEs) are nuclear transients powered by the disruption and subsequent accretion of stellar debris by a massive black hole \citep{Hills1975, Rees1988, Gezari2021}. They offer a direct way to study short-lived accretion episodes and the response of gas and dust in galactic nuclei. The observed emission can arise from physically distinct regions, ranging from the inner accretion flow, traced by X-ray and ultraviolet (UV) radiation, to the surrounding circumnuclear medium, traced by optical, infrared (IR), and radio emission \citep{Gezari2021, vanVelzen2021}.

Optical and soft X-ray surveys have historically dominated TDE discovery, but these bands are strongly affected by obscuration, viewing angle, and reprocessing geometry. Dust-obscured or IR-selected TDE-like events may therefore be
missed by optical and X-ray searches
\citep{Jiang2021MIRONG,Masterson2024,Yao2025}. In such systems, a substantial fraction of the primary UV/optical luminosity can be absorbed by circumnuclear dust and re-emitted at mid-infrared (MIR) wavelengths. The MIR dust echo can therefore provide a partial measure of the energy released by a hidden nuclear flare, although the inferred energy and scale depend on the dust sublimation radius, radial distribution, geometry, and covering factor \citep{Lu2016, vanVelzen2021}.

MIR echoes detected in optically selected TDEs have shown that \textit{WISE} W1 and W2 variability can constrain the effective dust temperature, luminosity, covering factor, and emitting scale \citep{vanVelzen2016, Jiang2016, Jiang2021}. Long-lived MIR declines have also been found in transient coronal-line emitters, non-Seyfert galaxies, and dusty nuclear transients, supporting the interpretation that some nuclear MIR flares are delayed dust echoes of energetic accretion events \citep{Dou2016, Wang2018, Reynolds2022}. Large-amplitude MIR flares without optical counterparts can also arise from obscured TDE-like events, turning-on AGN, or other nuclear accretion transients, which makes additional MIR diagnostics essential for interpreting their physical origin \citep{Yang2019}.

Independent evidence from time-series spectropolarimetry of the nearby nuclear transient AT2023clx further illustrates that TDE-like events can probe pre-existing dusty structures in galactic nuclei, including the possible geometrical connection between a transient outflow and the nuclear dusty region \citep{Uno2025}. Recent IR studies have also shown that the infrared response of TDEs can include viewing-angle effects, non-dust reprocessing at early times, and clumpy or extended dust structures at later times, so a single simple dust shell is not always an adequate description \citep{Reynolds2026, Wu2025_AT2019qiz}. In this context, hydrodynamical TDE fallback-rate libraries can provide physically motivated input light curves for consistency tests, while dust-echo calculations show that the observed MIR response depends on the dust geometry, temperature structure, covering factor, and light-travel-time delays \citep{LawSmith2020STARS, Lu2016}.

The WISE mission and its NEOWISE reactivation phase have provided long-term W1 and W2 light curves, enabling systematic searches for nuclear transients at MIR wavelengths \citep{Wright2010, Mainzer2014}. Among the earliest systematic searches for MIR-selected nuclear
transients, the Mid-Infrared Outbursts in Nearby Galaxies (MIRONG)
programme identified a large population of nuclear MIR flares
\citep{Jiang2021MIRONG}, with optical spectroscopic follow-up revealing
transient spectral signatures consistent with TDEs and other episodes
of transient SMBH accretion \citep{Wang2022}. Subsequent WISE-based
searches have expanded the population of MIR-selected TDE candidates
and dust-echo-like infrared flares, particularly among obscured or
optically weak events whose primary flare is hidden, poorly sampled,
or missed entirely \citep{Masterson2024,Necker2025}. The relative evolution of $W1$ and $W2$ provides additional constraints beyond the light-curve shape: colour changes, flux--flux behaviour, and internal $W2$--$W1$ timing can quantify the degree of phase-dependent complexity in the MIR response and search for evidence of a structured dust response \citep{Yao2025}.

The host-galaxy environment is an important factor in interpreting nuclear activity. Isolated galaxies provide a low-interaction environment in which recent external perturbations are reduced relative to galaxies in pairs, groups, or denser large-scale environments \citep{Kompaniiets2025Environment}. Previous studies have investigated the morphology and isolation properties of such systems \citep{Vavilova2009, Karachentseva2010}. Isolated AGNs selected from the 2MIG catalogue have been shown to exhibit diverse nuclear and host-galaxy properties while evolving in low-interaction environments \citep{Pulatova2015, Pulatova2023, Kompaniiets2023, Kompaniiets2025b, Izviekova2026}. This environmental context makes UGC~11487 a relevant target for studying a MIR-selected TDE-like nuclear flare in an isolated barred galaxy.

UGC~11487, also known as WTP14adeqka, was identified by the WISE Transient Pipeline as a MIR-selected TDE candidate \citep{Masterson2024} and is included among the brightest dust-echo-like infrared flares in the Flaires catalogue \citep{Necker2025}. It was subsequently followed up at radio wavelengths, where delayed radio emission and a resolved sub-parsec outflow were reported \citep{Golay2026}. Together, the MIR and radio observations provide complementary probes of the compact nuclear environment, with WISE tracing thermal dust reprocessing and the radio emission tracing synchrotron-emitting plasma associated with an outflow or shock. However, the internal W1--W2 evolution, the phase-dependent MIR behaviour, and their connection to the compact radio-emitting region have not yet been analyzed in detail.

Because the available optical coverage does not densely sample the primary flare, a classical optical--MIR reverberation measurement is not possible. We found no corresponding optical transient in ASAS-SN or ATLAS \citep{Kochanek2017, Tonry2018}; this non-detection is not strongly constraining, because the optical counterpart could have been faint, obscured, or below the effective depth of these wide-field surveys. The absence of a well-sampled optical driver motivates an analysis based on the internal WISE W1--W2 evolution.

The aim of this work is to characterize the MIR evolution of UGC~11487 and to test whether phase-resolved W1--W2 diagnostics can reveal a structured multi-zone circumnuclear dust response in an obscured nuclear transient. We combine empirical timing diagnostics, phenomenological decomposition, and axisymmetric dust radiative-transfer models driven by stellar fallback histories to investigate the delayed W2 emission and compare the resulting MIR scales with independent radio constraints.

The paper is organized as follows. Section~\ref{sec:target_data} describes the WISE/NEOWISE data, ancillary optical information, and the published radio constraints. Section~\ref{sec:methods} presents the light-curve preparation, empirical $W1$--$W2$ diagnostics, effective colour-blackbody analysis, and  phenomenological and radiative-transfer modeling. Section~\ref{sec:results} summarizes the MIR timing, colour, energetics, and model-comparison results. Section~\ref{sec:discussion} discusses the interpretation of UGC~11487 as a structured dust response to an obscured TDE-like nuclear flare and compares the MIR-derived scales with the radio-emitting region. The main conclusions are given in Section~\ref{sec:conclusions}. Appendices~\ref{app:phenomenological_model_details}--\ref{app:CIGALE_input} provide the technical definitions, model details, auxiliary checks, robustness tests, and supplementary host-galaxy SED analysis.

\section{Target, observational data, and external constraints}
\label{sec:target_data}

UGC~11487 hosts the nuclear MIR transient WTP14adeqka, which was identified
as a MIR-selected TDE candidate by the WISE Transient Pipeline
\citep{Masterson2024}. It is also listed among the brightest
dust-echo-like IR flares in the Flaires catalogue \citep{Necker2025}. The
adopted MIR position of WTP14adeqka is
\(\alpha_{\rm J2000}=19^{\rm h}49^{\rm m}24\fs86\) and
\(\delta_{\rm J2000}=+63^\circ30^\prime33\farcs4\). The main properties of
the target and the external constraints used in this work are summarized in
Table~\ref{tab:target_properties}.

Independent radio observations from \citet{Golay2026} provide external
constraints on the compact nuclear environment.
We use the published radio luminosity, radio-SED-derived
equipartition energy and radius, VLBA-resolved sizes, and expansion
velocities as external reference quantities for comparison with the
MIR-derived dust-response scales. These radio constraints are summarized
in Table~\ref{tab:target_properties}. Together, the WISE and radio datasets
allow us to compare the characteristic scales of thermal dust reprocessing
and compact radio-emitting material in the nucleus of UGC~11487.

The main dataset used in this work is the WISE/NEOWISE time-domain
photometry in W1 and W2, centred at 3.4 and 4.6~\(\mu{\rm m}\),
respectively \citep{Wright2010,Mainzer2014}. These data are used to trace the MIR evolution of the flare, measure the internal $W1$--$W2$ timing and colour evolution, and assess the phase dependence and characteristic complexity of the MIR dust response.

We also use ZTF \(gri\) photometry as ancillary optical data
\citep{Bellm2019}. Since these observations do not cover the primary MIR
rise and peak, they are used only to check for optical variability during the
MIR decline. The auxiliary ZTF colour and flux--flux checks are summarized
in Appendix~\ref{app:ztf}.

We further use archival multiwavelength photometry to construct an
aperture-matched integrated host-galaxy SED of UGC~11487. This SED is used
to characterize the time-averaged stellar, star-forming, and dust content of
the host galaxy. The aperture photometry and SED-modeling procedure are
described in Sect.~\ref{subsec:sed_photometry_methods}.

\begin{table}
\centering
\caption{Literature parameters of UGC~11487 / WTP14adeqka and external constraints used in this work.}
\label{tab:target_properties}
\setlength{\tabcolsep}{3.5pt}
\renewcommand{\arraystretch}{1.12}

\begin{tabular}{l c c}
\hline
Quantity & Value & Ref. \\
\hline
Redshift $z$ & 0.018950 & [5] \\
$D_{\rm L}$ & 82.34 Mpc & [5] \\
Environment & Isolated, barred & [2] \\
Transient class & MIR-selected TDE cand. & [3] \\
Flaires class & Bright dust-echo flare & [4] \\
$\log(M_{\rm SMBH}/M_\odot)$ & $6.8\pm0.5$ & [3] \\
$R_{\rm dust,shell}$ & $0.33^{+0.13}_{-0.05}$ pc & [3] \\[1.2ex]
\hline

Peak radio $\nu L_\nu$ &
\rule{0pt}{2.6ex}$\simeq2\times10^{39}$ erg s$^{-1}$ & [1] \\

{Radio $E_{\rm eq}$} &
{ $\simeq10^{49.8}$ erg} &
[1] \\

{Radio SED $R_{\rm eq}$} &
{ $0.060$--$0.065$ pc} &
[1] \\

{SED-based $\beta$} &
{ $0.022$--$0.031$} &
[1] \\

{VLBA $\sigma_{\rm maj}$} &
{ $0.111$--$0.127$ pc} &
[1] \\

{VLBA major-axis $\beta$} &
{ $0.049$} &
[1] \\

{VLBA PA-averaged $\beta$} &
{ $0.025$} &
[1] \\[0.6ex]

\hline
\end{tabular}

\tablefoot{
References: [1] \citet{Golay2026}; [2] \citet{Pulatova2015};
[3] \citet{Masterson2024}; [4] \citet{Necker2025};
[5] NASA/IPAC Extragalactic Database (NED).
The radio SED radii are the cooling-break-corrected values
$R_{\rm eq}=0.065\pm0.003$ and $0.060\pm0.003$ pc for the two
multifrequency epochs, with $E_{\rm eq}\simeq10^{49.8}$ erg.
The SED-based velocity range corresponds to launch delays of
0--1000 d. The VLBA sizes are the major-axis $\sigma_{\rm maj}$
scales of the fitted elliptical Gaussian models. The corresponding
major-axis and position-angle-averaged expansion velocities are
$\beta\simeq0.049$ and $\beta\simeq0.025$, respectively
\citep{Golay2026}.
}
\end{table}

\section{Methods}
\label{sec:methods}

\subsection{WISE/NEOWISE photometry and baseline}
\label{subsec:wise_photometry}

We analysed epoch-averaged WISE/NEOWISE $W1$ and $W2$ photometry of UGC~11487 from 2014 to 2024. To constrain the pre-flare MIR level, we also queried the AllWISE multi-epoch photometry table (\texttt{allwise\_p3as\_mep}), which provides two 2010 $W1/W2$ epoch-averaged measurements. Together with the first pre-rise NEOWISE epochs, these data define the quiescent MIR baseline adopted for the baseline-subtracted excess light curves.

Single-exposure NEOWISE measurements were retained when both $W1$ and $W2$ had valid photometry, $S/N(W1)>5$, $S/N(W2)>3$, $\texttt{cc\_flags}=00$, and $\texttt{qual\_frame}>0$. The cleaned measurements were grouped according to the natural NEOWISE cadence. The retained NEOWISE measurements have stable PSF-fit diagnostics, $\texttt{w1fitr}=\texttt{w2fitr}=7.5\arcsec$, and no saturation flags, $\texttt{w1sat}=\texttt{w2sat}=0$.

The fiducial pre-flare baseline was defined as the median flux density of the two 2010 AllWISE epochs and the first two pre-rise NEOWISE epochs, yielding $F_{W1,{\rm base}}=8.453~{\rm mJy}$ and $F_{W2,{\rm base}}=5.428~{\rm mJy}$. The adopted baseline combines the earliest AllWISE and pre-rise NEOWISE measurements to reduce sensitivity to possible low-level variability and to provide a robust estimate of the quiescent MIR emission. The 2010 points were used only to define the baseline and were excluded from the flare fluence and energy integration.

WISE Vega magnitudes were converted to flux densities as
$F_\nu = F_0\,10^{-0.4m}$.
We adopted the WISE Vega zero-magnitude flux densities
$F_{0,W1}=309.54~{\rm Jy}$ and $F_{0,W2}=171.787~{\rm Jy}$
from the WISE All-Sky Data Release Explanatory Supplement
\citep{Cutri2012WISE}. The baseline-subtracted excess flux density was defined as $F_\nu(t)=F_\nu(t)-F_{\nu,{\rm base}} $
The timing, fluence, and colour-blackbody diagnostics use these baseline-subtracted excess light curves. The colour--magnitude relation in Fig.~\ref{fig:wise_colour_magnitude} instead uses the observed total-flux Vega magnitudes.

For model fitting and Monte Carlo resampling, we adopted an effective flux-density uncertainty
\begin{equation}
\sigma_{\rm eff} =
\left(\sigma_{F_\nu}^{2}+\sigma_{\rm floor}^{2}\right)^{1/2},
\label{eq:effective_error}
\end{equation}
where $\sigma_{F_\nu}$ is the formal WISE flux-density uncertainty and
$\sigma_{\rm floor}=1.10~{\rm mJy}$.
The fiducial value follows from the effective-error prescription
\[
\sigma_{\rm floor}
=
\max\left[
0.30~{\rm mJy},
\,0.03\max\left(|\Delta F_{W1}|,|\Delta F_{W2}|\right)
\right],
\]
which gives $1.09~{\rm mJy}$ for the maximum MIR excess of
$36.47~{\rm mJy}$ in UGC~11487. The resulting $1.10~{\rm mJy}$ additive
term was applied throughout the fitting and Monte Carlo calculations to
limit the weight of very small formal epoch uncertainties.

As a robustness test, we repeated the one-, two-, and
three-component model comparison using the flux-dependent prescription
\[
\sigma_{\rm eff,10\%}
=
\left[
\sigma_{F_\nu}^{2}
+
(0.10 F_\nu)^{2}
\right]^{1/2},
\]
motivated by \citet{Masterson2024}, where $F_\nu$ is the observed
total epoch flux density. The corresponding model-selection results are
given in Appendix~\ref{app:baseline_sensitivity}
(Table~\ref{tab:error_model_sensitivity}).

The uncertainty of the adopted pre-flare baseline was included separately
in the Monte Carlo procedure by bootstrap resampling of the baseline epochs
and perturbing their fluxes within their formal photometric uncertainties.

\subsection{Empirical \texorpdfstring{$W1$--$W2$}{W1--W2} diagnostics}
\label{subsec:empirical_diagnostics}

To characterize the internal phase-dependent MIR evolution, we used empirical $W1-W2$ diagnostics based on colour evolution, colour--magnitude behaviour, flux--flux relations, and model-independent timing measures. This set of diagnostics traces both the spectral and temporal evolution of the MIR flare and quantifies the degree of phase-dependent complexity in the WISE response.

Observed-frame epochs were converted to rest-frame time relative to the first NEOWISE epoch,
\begin{equation}
t_{\rm rest}=\frac{t_{\rm obs}-t_{\rm ref}}{1+z}.
\label{eq:rest_time}
\end{equation}

The WISE colour index $W1-W2$ and the colour--magnitude relation were used to follow the MIR spectral evolution. The colour--magnitude relation was fitted using the York regression \citep{York2004}, accounting for correlated uncertainties between $W1-W2$ and $W1$. Flux--flux fits were performed with orthogonal distance regression
\citep{Boggs1990} for both absolute and baseline-subtracted flux densities.

The rise and decline occupy different loci in the flux--flux plane. For this reason, we fitted these phases separately to quantify potential hysteresis and deviations from a single-component evolution. We applied both a broad rise--decline split and a stricter division into the clean rise, turnover, and post-peak decline phases (Fig.~\ref{fig:wise_fluxflux_hysteresis}).

In the absence of a well-sampled optical or UV driver, classical reverberation mapping cannot be applied. We therefore used internal MIR timing diagnostics based solely on the relative evolution of $W1$ and $W2$. The $W2$--$W1$ peak-time offset was defined as
\begin{equation}
\Delta t_{\rm peak}=t_{\rm peak}(W2)-t_{\rm peak}(W1),
\label{eq:peak_offset}
\end{equation}
and the positive-fluence centroid offset was defined as
\begin{equation}
\Delta t_{\rm cent}=t_{\rm cent}(W2)-t_{\rm cent}(W1),
\label{eq:centroid_offset}
\end{equation}
where
\begin{equation}
t_{\rm cent}=
\frac{\int t\,\max[\Delta F(t),0]\,dt}
{\int\max[\Delta F(t),0]\,dt}.
\label{eq:fluence_centroid}
\end{equation}
The corresponding positive MIR fluence in each WISE band was
\begin{equation}
\Phi_{Wj}=
\int\max[\Delta F_{Wj}(t),0]\,dt,
\qquad j=1,2.
\label{eq:positive_fluence}
\end{equation}
We characterized the temporal extent of the positive excess by its
flux-weighted root-mean-square width about the positive-fluence centroid,
\begingroup

\begin{equation}
\begin{aligned}
\sigma_{t,Wj}
&=
\left[
\frac{\sum_i w_{j,i}(t_i-t_{{\rm cent},Wj})^2}
{\sum_i w_{j,i}}
\right]^{1/2},\\
w_{j,i}
&=\max[\Delta F_{Wj}(t_i),0].
\end{aligned}
\label{eq:positive_width}
\end{equation}
\endgroup
Here $t_i$ are the rest-frame epoch times, so only epochs with
positive baseline-subtracted excess contribute to $\sigma_{t,Wj}$.

The integrals were evaluated numerically from the epoch-averaged light curves using trapezoidal integration.
The temporal widths were evaluated directly from the positive
epoch excesses using Eq.~\ref{eq:positive_width}. The 2010 AllWISE points were used only to define and resample the pre-flare baseline and were excluded from the flare fluence and energy integrations.

The positive-fluence centroid was adopted as the preferred timing measure because it incorporates the full positive-excess light curve. This reduces the dependence on a single sampled maximum and makes the estimate more robust against the sparse WISE cadence.

The corresponding timing offsets were expressed as $c\Delta t_{\rm rest}$ to place the internal $W1$--$W2$ timing difference on an approximate light-travel-distance scale. This quantity represents a characteristic separation between the $W1$- and $W2$-weighted dust responses. It is distinct from an absolute dust radius or an optical--MIR reverberation radius. The sensitivity of the adopted timing diagnostics to the baseline definition is tested in Appendix~\ref{app:baseline_sensitivity}.

We also used a shift-and-broaden $W1$-to-$W2$ template test to characterize the shape difference between the two MIR bands. In this test, the $W2$ excess light curve was represented as an amplitude-scaled, time-shifted, and temporally smoothed version of the $W1$ excess light curve,
\begin{equation}
\Delta F_{W2}(t)
=
A\,
\left[
G_s \ast \Delta F_{W1}
\right](t-\tau),
\label{eq:shift_broaden}
\end{equation}
where $A$ is an amplitude scale factor, $\tau$ is the compact $W1$-to-$W2$ shift, $G_s$ is a Gaussian smoothing kernel with smoothing timescale $s$, and $\ast$ denotes convolution. The fitted shift, $\tau$, describes the compact $W1$-to-$W2$ displacement after allowing for additional smoothing of the $W2$ response, while the smoothing timescale captures the broader temporal width of $W2$. This diagnostic separates the compact $W1$-to-$W2$ displacement from the additional temporal broadening of the $W2$ response. The fit was performed on a grid of $\tau$ and $s$, with the optimal amplitude obtained by weighted least squares for each grid point.

Uncertainties for the empirical diagnostics in this subsection were estimated through Monte Carlo resampling of the WISE fluxes, including bootstrap resampling of the pre-flare baseline.

Together, the colour, flux--flux, timing, and shift-and-broaden diagnostics probe different aspects of the MIR evolution. They are complemented by effective colour-blackbody quantities, which provide an empirical description of the temperature and emitting-scale evolution of the MIR flare.

\subsection{Effective colour-blackbody diagnostics}
\label{subsec:colour_blackbody_method}

We estimated effective two-band colour-blackbody quantities from the baseline-subtracted $W1$ and $W2$ excess fluxes, $\Delta F_W=F_W-F_{W,{\rm base}}$. These quantities provide empirical diagnostics of the temperature, emitting scale, and luminosity evolution of the MIR flare. They should be interpreted as descriptive thermal parameters of the WISE excess emission.

For epochs with $\Delta F_{W1}>0$ and $\Delta F_{W2}>0$, the effective colour temperature was obtained from the ratio of the $W1$ and $W2$ excess flux densities,
\begin{equation}
\frac{\Delta F_{W1}}{\Delta F_{W2}}
=
\frac{B_\nu(\nu_{W1,{\rm rest}},T_{\rm eff})}
{B_\nu(\nu_{W2,{\rm rest}},T_{\rm eff})},
\label{eq:teff_ratio}
\end{equation}
where $\nu_{\rm rest}=(1+z)\nu_{\rm obs}$. After $T_{\rm eff}$ was determined, the effective blackbody radius was estimated from the $W1$ excess flux density,
\begin{equation}
\Delta F_{\nu,{\rm obs}}
=
\frac{(1+z)\pi R_{\rm BB}^{2}
B_\nu(\nu_{\rm rest},T_{\rm eff})}
{D_{\rm L}^{2}} .
\label{eq:rbb}
\end{equation}
The approximate bolometric IR luminosity was then calculated as
\begin{equation}
L_{\rm IR}
=
4\pi R_{\rm BB}^{2}\sigma_{\rm SB}T_{\rm eff}^{4},
\label{eq:lir}
\end{equation}
and the integrated effective IR energy as
\begin{equation}
E_{\rm IR}
=
\int L_{\rm IR}(t)\,dt ,
\label{eq:eir}
\end{equation}
with the integration performed over rest-frame time. Uncertainties were evaluated through Monte Carlo resampling of both the WISE fluxes and the pre-flare baseline.

An epoch was classified as reliable if it was not used in the baseline definition, both $W1$ and $W2$ excess fluxes satisfied $\Delta F/\sigma_{\rm eff}\geq2$, and the resulting Monte Carlo distributions were sufficiently constrained. Specifically, we required fractional uncertainties below 0.35 for $T_{\rm eff}$ and below 0.75 for both $R_{\rm BB}$ and $L_{\rm IR}$. Finite colour-blackbody solutions that did not satisfy these criteria were retained separately as poorly constrained estimates.

The derived blackbody radii represent effective emitting scales. They depend on the two-band colour temperature, the adopted baseline, and the assumption of a blackbody-like MIR excess. For this reason, they provide empirical constraints on the thermal evolution of the flare, while the underlying dust geometry remains degenerate.

\rev{The effective IR energy gives an illustrative accreted-mass scale}
\begingroup
\begin{equation}
M_{\rm acc}
=
\frac{E_{\rm IR}}{f_{\rm cov}\eta_{\rm rad}c^2},
\label{eq:macc}
\end{equation}
\endgroup
\rev{where $f_{\rm cov}$ is the effective dust reprocessing factor and $\eta_{\rm rad}$ is the radiative efficiency.}
The corresponding disrupted-stellar mass scale was estimated as
\begin{equation}
M_{\star,\rm scale}
=
\frac{M_{\rm acc}}{f_{\rm acc}},
\label{eq:mstar_scale}
\end{equation}
where $f_{\rm acc}$ is the fraction of the disrupted stellar mass that is ultimately accreted. These quantities provide order-of-magnitude energetic consistency estimates and depend on the adopted values of $f_{\rm cov}$, \rev{$\eta_{\rm rad}$}, and $f_{\rm acc}$.

\rev{These effective quantities provide complementary diagnostics of the same two-band data used to test the structured dust-response scenario.}

\subsection{Phenomenological decomposition and physical
dust-echo modeling}
\label{subsec:model_methods}

We fitted the $W1$ and $W2$ excess light curves simultaneously with one-, two-, and three-component phenomenological models. Each component was represented by a smooth, asymmetric temporal profile, with the $W1$ and $W2$ amplitudes linked by the Planck ratio evaluated at the effective rest-frame WISE frequencies and a constant component colour temperature
$T_i$. The component definitions are given in Appendix~\ref{app:phenomenological_model_details}. The models were compared using the Akaike information criterion (AIC) and the Bayesian information criterion (BIC) \citep{Akaike1974, Schwarz1978}.

\begin{revisionblock}
We modelled compact and extended graphite populations in an effective smooth axisymmetric medium, illuminated by STARS fallback histories \citep{LawSmith2020STARS,LawSmith2020STARSZenodo} with causal accretion smoothing. Radiative equilibrium includes attenuated direct illumination and secondary dust heating; the observer calculation includes light-travel delays and wavelength-dependent attenuation. Synthetic $W1$ and $W2$ flux densities were obtained using the WISE equal-energy relative spectral response curves and Vega-system zero points \citep{Jarrett2011},
with the CALSPEC Vega reference spectrum \texttt{alpha\_lyr\_stis\_011.fits} \citep{Bohlin2020}.
\end{revisionblock}

The physical calculation follows the sequence
\begin{equation}
\begin{gathered}
\dot M_{\rm fb}(t)
\longrightarrow
\dot M_{\rm acc}(t)
\longrightarrow
L_{\rm bol}(t)
\\
\longrightarrow
T(r,\theta,a,u)
\longrightarrow
F_{W1,W2}^{\rm model}(t_{\rm obs}),
\end{gathered}
\label{eq:physical_model_sequence}
\end{equation}

\begin{revisionblock}
Here $u$ is source-retarded time, and the fluxes include the transmitted source and fixed host emission. Appendix~\ref{app:hydro_clumpy_details} specifies the physical assumptions and validation; implementation details are in the \href{https://github.com/OlenaKompaniiets/tde-dust-echo/tree/b459f1f99ad38a42db59d798062b344c84343ab8}{version-pinned code}. More detailed description can be found on the \href{https://olenakompaniiets.github.io/tde-dust-echo/index.html}{model page at GitHub}. 
\end{revisionblock}

\begin{revisionblock}
The physical models were fitted jointly to 22 NEOWISE epochs in each band using the effective uncertainties of Eq.~\ref{eq:effective_error}. For each fixed STARS history and radial configuration, seven continuous parameters were adjusted: $\eta$, $\tau_{\rm visc}$, $\tau_{\rm C}$, $\tau_{\rm E}$, the opening angle, inclination, and event epoch $t_0$. Thus, $N=44$ gives the nominal conditional $\nu=N-k=37$. Host flux densities and grain properties were fixed.
The discrete source and geometry selection, approximate conditional parameter variations, and direct-solver checks of these variations are described in Appendix~\ref{app:hydro_clumpy_details}.
\end{revisionblock}.

\subsection{Robustness checks}
\label{subsec:robustness_consistency}

We tested the stability of the $W2-W1$ centroid scale against three baseline definitions: pre-rise NEOWISE only, the fiducial AllWISE+NEOWISE baseline, and AllWISE 2010 only. We also tested the dependence of the smooth component comparison on the adopted effective error floor.

The possible late-time $W2$ residual around epochs 17--19, marked in the numbered WISE light curve in Fig.~\ref{fig:wise_lightcurve}, was assessed by comparing those epochs with a smooth post-peak decline fitted after excluding them. This test evaluates the significance of possible deviations from a smooth decline and treats the late-time $W2$ excess as a residual feature whose interpretation remains model-dependent. Details of these tests are given in Appendix~\ref{app:baseline_sensitivity}.

\subsection{Host-galaxy environment from UV-to-radio SED modeling}
\label{subsec:sed_photometry_methods}

\rev{We constructed an aperture-matched UV-to-radio spectral energy distribution to characterize the non-transient host-galaxy emission, assuming an absent or weak persistent AGN component ($0\leq f_{\rm AGN}\leq0.1$).}

We used archival data for UGC~11487 from GALEX
\citep{Morrissey2007}, Pan-STARRS1 \citep{Chambers2016},
2MASS \citep{Skrutskie2006}, WISE \citep{Wright2010,Cutri2012WISE},
IRAS/IRIS \citep{Neugebauer1984,MivilleDeschenes2005}, and
NVSS \citep{Condon1998}. For the photometric measurements, we matched full-galaxy apertures where angular resolution allowed, with separate low-resolution apertures for WISE W3--W4, IRAS, and NVSS. The resulting broadband fluxes were modeled with CIGALE to estimate the time-averaged stellar mass, star formation rate, attenuation, and dust luminosity. The detailed aperture definitions, masking procedure, photometric conversions, treatment of IRAS upper limits, and validation images are given in Appendix~\ref{app:host_sed_methods}.

\rev{The host-galaxy SED modeling provides the environmental context for the MIR flare, complementing the phase-resolved diagnostics.}

\section{Results}
\label{sec:results}

The main empirical quantities derived from the WISE/NEOWISE analysis are summarized in Table~\ref{tab:mir_summary}. The model-independent $W1$--$W2$ timing diagnostics are listed in Table~\ref{tab:mir_timing_summary}, while the effective colour-blackbody quantities are summarized in Table~\ref{tab:teff_summary}. The smooth phenomenological model comparison is given in Table~\ref{tab:model_comparison}. The MIR--radio scale comparison is illustrated in Fig.~\ref{fig:mir_radio_scales}, with the corresponding illustrative travel-time estimates given in Appendix~\ref{app:baseline_sensitivity}.

\subsection{Phase-dependent WISE morphology and \texorpdfstring{$W1$--$W2$}{W1--W2} hysteresis}
\label{subsec:wise_results}
\label{subsec:colour_results}

\begin{figure}
\centering
\includegraphics[width=\columnwidth]{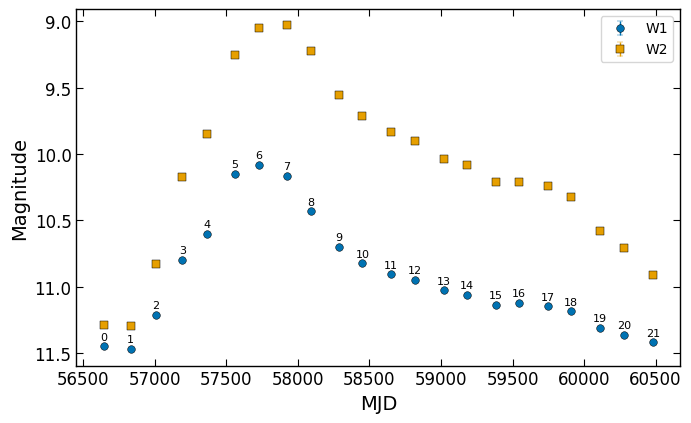}
\caption{
Epoch-averaged WISE/NEOWISE $W1$ and $W2$ light curves of UGC~11487. The source shows a decade-long MIR flare with a larger peak-to-peak magnitude amplitude in $W2$ than in $W1$. The numbered labels mark the WISE epochs used throughout the analysis.
}
\label{fig:wise_lightcurve}
\end{figure}

The epoch-averaged $W1$ and $W2$ light curves of UGC~11487 show a broad,
asymmetric MIR flare lasting nearly a decade
(Fig.~\ref{fig:wise_lightcurve}). The peak-to-peak magnitude amplitude
measured from the epoch-averaged light curves is larger in $W2$,
$A_{W2}\simeq2.27$ mag, than in $W1$, $A_{W1}\simeq1.39$ mag.
The time-integrated excess is also larger in $W2$: integrating
the baseline-subtracted positive excess over the flare gives a
$W2$-to-$W1$ positive-fluence ratio of $2.23\pm0.12$
(Table~\ref{tab:mir_summary}). The positive-excess temporal width is
$\simeq781$ d in $W2$, compared with $\simeq697$ d in $W1$, independently
showing that the $W2$ response is broader in time. These results show
chromatic MIR variability with a substantial contribution from cooler
dust traced most strongly by $W2$.

The peak of the excess emission occurs first in $W1$ and later in $W2$: the $W1$ excess reaches its maximum at epoch 6, whereas the $W2$ excess peaks at epoch 7 (Fig.~\ref{fig:wise_lightcurve}). The $W2$ light curve also shows a slower decline and remains above the baseline after the $W1$ excess has substantially faded. This morphology already suggests that the MIR evolution is phase dependent and motivates the internal $W1$--$W2$ timing analysis presented below.

\begin{figure}
\centering
\includegraphics[width=\columnwidth]{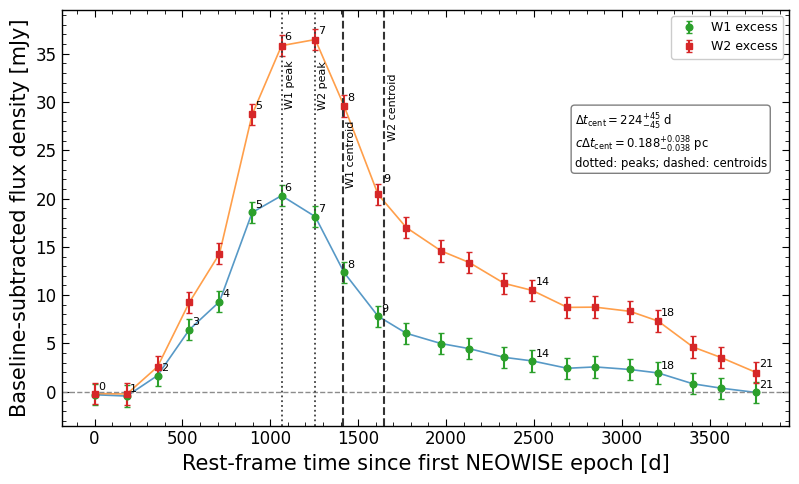}
\caption{
Baseline-subtracted WISE $W1$ and $W2$ excess light curves of UGC~11487 using the combined AllWISE 2010 plus pre-rise NEOWISE baseline. Vertical dotted lines mark the $W1$ and $W2$ peak epochs, while dashed lines show the positive-fluence centroids. $W2$ peaks later and declines more slowly than $W1$, indicating a broader cooler-dust response.
}
\label{fig:wise_excess_timing}
\end{figure}

\begin{figure}
\centering
\includegraphics[width=\columnwidth]{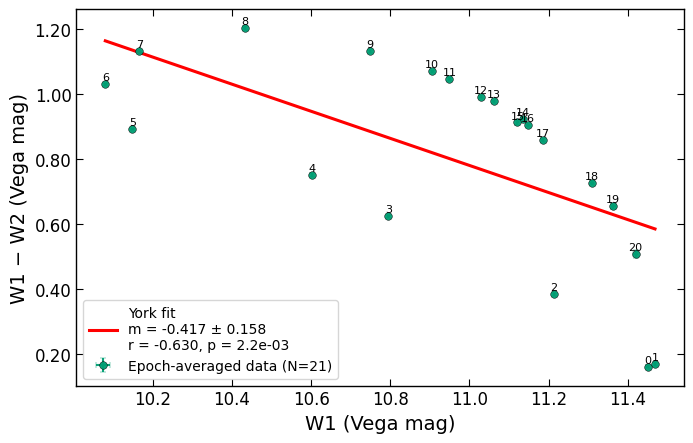}
\caption{
\rev{Observed-total-flux WISE $W1-W2$ versus $W1$ colour--magnitude relation for UGC~11487. The negative York slope indicates redder-when-brighter behaviour of the combined host and transient emission.}
}
\label{fig:wise_colour_magnitude}
\end{figure}

\rev{The observed-total-flux WISE $W1-W2$ versus $W1$ colour--magnitude relation for UGC~11487 shows a redder-when-brighter (RWB) trend} (Fig.~\ref{fig:wise_colour_magnitude}). The York fit gives a slope of $-0.42\pm0.16$, with a Pearson correlation coefficient of approximately $-0.63$. \rev{This total-flux trend includes the changing contrast between the host and transient emission; it is distinct from the excess-only colour-temperature evolution.}

\begin{figure}
\centering
\includegraphics[width=\columnwidth]{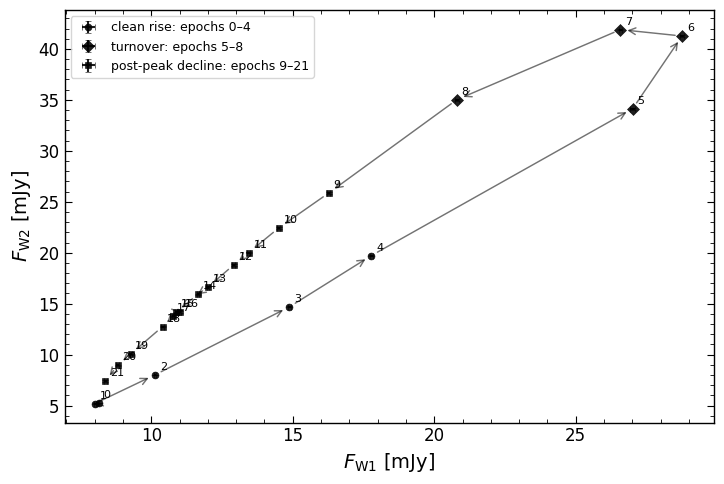}
\caption{
WISE flux--flux trajectory of UGC~11487. Arrows indicate the temporal direction of the MIR evolution. The source follows different paths during the rise and decline: at a similar $F_{W1}$, the declining branch has a larger $F_{W2}$. This hysteresis-like trajectory indicates that the cooler $W2$-emitting component remains bright after the hotter $W1$-emitting component has faded.
}
\label{fig:wise_fluxflux_hysteresis}
\end{figure}

The phase dependence is more apparent in the WISE flux--flux trajectory (Fig.~\ref{fig:wise_fluxflux_hysteresis}). During the rise, the source follows a lower branch in the $F_{W2}$--$F_{W1}$ plane. During the decline, it returns along an upper branch, with larger $F_{W2}$ at a similar $F_{W1}$. This hysteresis-like trajectory demonstrates that the MIR spectral evolution depends on phase and suggests a delayed and more persistent $W2$-emitting component.

The phase-resolved flux--flux fits quantify this behaviour. For the broad rise--decline division, the rising branch, defined by epochs 0--6, has $m=1.51\pm0.09$, whereas the declining branch, defined by epochs 7--21, has $m=2.01\pm0.18$. A stricter split excluding the turnover region, epochs 5--8, gives $m_{\rm rise}=1.43\pm0.12$ for epochs 0--4 and $m_{\rm decline}=2.34\pm0.06$ for epochs 9--21. The declining branch is therefore substantially more $W2$-enhanced than the rising branch, quantitatively confirming the phase dependence of the MIR evolution. Together, these empirical diagnostics show that the MIR evolution of UGC~11487 is phase dependent and departs from a self-similar single-component trajectory, motivating the timing analysis presented in Sect.~\ref{subsec:timing_results}.

\begin{table*}
\centering
\caption{Summary of the empirical MIR properties derived for UGC~11487.
The positive-excess temporal width is the flux-weighted RMS width
defined in Eq.~\ref{eq:positive_width}.}
\label{tab:mir_summary}
\begin{tabular}{l l l}
\hline
Quantity & Value & Interpretation \\
\hline
WISE/NEOWISE coverage & 22 epoch-averaged points & $W1/W2$ MIR light curve, 2014--2024 \\
Pre-flare baseline & 2 AllWISE + 2 NEOWISE epochs & Combined low-state baseline \\
MIR event duration & $\sim10$ yr & Long-lived nuclear MIR flare \\
$A_{W1}$ & $\simeq1.39$ mag & Peak-to-peak $W1$ amplitude \\
$A_{W2}$ & $\simeq2.27$ mag & Peak-to-peak $W2$ amplitude \\
Median $W1-W2$ & $\simeq0.91$ mag & Red MIR colour during the event \\
Colour--magnitude slope & $-0.42\pm0.16$ & Redder when brighter in MIR \\
Global flux--flux slope & $2.17\pm0.21$ & Overall $W2$-enhanced variability \\
Rise flux--flux slope & $1.51\pm0.09$ & Rising branch, epochs 0--6 \\
Decline flux--flux slope & $2.01\pm0.18$ & Declining branch, epochs 7--21 \\
$W1$ positive-excess width
& $\simeq697$ d
&Flux-weighted RMS width of the positive excess \\
$W2$ positive-excess width
& $\simeq781$ d
& Flux-weighted RMS width of the positive excess \\
$W1$ positive fluence & $\simeq2.29\times10^4$ mJy d & Integrated $W1$ excess \\
$W2$ positive fluence & $\simeq5.12\times10^4$ mJy d & $W2$-dominated MIR fluence \\
$W2/W1$ fluence ratio & $2.23\pm0.12$ & $W2$ carries about twice the positive fluence \\
\hline
\end{tabular}
\end{table*}

\begin{table*}
\centering
\caption{
Model-independent $W1$--$W2$ timing and scale diagnostics. }
\label{tab:mir_timing_summary}
\begin{tabular}{l c c l}
\hline
Diagnostic & Value & Light-travel scale & Comment \\
\hline
$W1$ excess peak & epoch 6, MJD 57727.5 & -- & Hotter $W1$-dominated maximum \\
$W2$ excess peak & epoch 7, MJD 57923.0 & -- & $W2$ maximum occurs later \\
$W2$--$W1$ peak offset & $191.8$ d & $0.161$ pc & Sensitive to sparse NEOWISE cadence \\
$W1$ positive-fluence centroid & ${\rm MJD}\ 58088.6\pm44.0$ & -- & Centroid of positive $W1$ excess \\
$W2$ positive-fluence centroid & ${\rm MJD}\ 58316.7\pm18.8$ & -- & Centroid of positive $W2$ excess \\
$W2$--$W1$ centroid offset
& $224\pm45$ d
& $0.188\pm0.038$ pc
& Effective differential $W1$--$W2$ response scale \\
\hline
\end{tabular}
\tablefoot{The peak and centroid offsets are given in the rest frame. The centroid values are positive-fluence centroids of the baseline-subtracted WISE excess light curves. The uncertainties include Monte Carlo resampling of the epoch fluxes and bootstrap resampling of the four pre-flare baseline candidates.}
\end{table*}

\begin{table*}
\centering
\caption{Effective $W1$--$W2$ colour-blackbody diagnostics.}
\label{tab:teff_summary}
\begin{tabular}{l c c c l}
\hline
Stage & $T_{\rm eff}$ & $R_{\rm BB}$ & $L_{\rm IR}$ & Interpretation \\
 & K & pc & erg s$^{-1}$ & \\
\hline
Rise & $\sim900$ & $\sim0.03$--$0.05$ & $\sim10^{43}$ & Hotter compact MIR-emitting region \\
Turnover/peak & $\sim700$--$850$ & $\sim0.10$--$0.15$ & $\sim(2$--$3)\times10^{43}$ & Maximum effective emitting area \\
Post-peak decline & $\sim500$--$600$ & $\sim0.10$--$0.15$ & $\sim10^{43}$ & Cooler longer-lived $W2$-dominated component \\
\hline
Integrated $E_{\rm IR}$, reliable epochs & \multicolumn{3}{c}{$\simeq2.9\times10^{51}$ erg} & Fiducial reliable-epoch estimate \\
Integrated $E_{\rm IR}$, all finite & \multicolumn{3}{c}{$\simeq3.8\times10^{51}$ erg} & All finite non-baseline estimates \\
\hline
\end{tabular}
\end{table*}

\begin{table}
\caption{Comparison of the one-, two-, and three-component smooth joint $W1+W2$ phenomenological models using the effective error floor.}
\label{tab:model_comparison}
\resizebox{\columnwidth}{!}{%
\begin{tabular}{l c c c c c c}
\hline
Model & $k$ & dof & $\chi^2_{\rm eff}$ & AIC$_{\rm eff}$ & BIC$_{\rm eff}$ & $\Delta{\rm BIC}$ \\
\hline
One component & 5  & 39 & 321.70 & 331.70 & 340.62 & 281.7 \\
Two components & 10 & 34 & 21.06  & 41.06  & 58.91  & 0.0 \\
Three components & 15 & 29 & 14.80 & 44.80  & 71.56  & 12.7 \\
\hline
\end{tabular}%
}
\tablefoot{These quantities are derived from the two-band WISE excess fluxes and provide effective diagnostics of the temperature, emitting scale, and luminosity evolution of the MIR flare.}
\end{table}

\subsection{Model-independent \texorpdfstring{$W2$--$W1$}{W2--W1} timing and effective MIR scale}
\label{subsec:timing_results}

\rev{The peak-based and positive-fluence centroid diagnostics both indicate delayed $W2$ emission relative to $W1$.} The $W2$ excess peak occurs $191.8$ d after the $W1$ excess peak, corresponding to $c\Delta t_{\rm peak}=0.161$ pc (Table~\ref{tab:mir_timing_summary}). This peak-based estimate is sensitive to the discrete NEOWISE sampling and is therefore used only as a supporting timing diagnostic.

The positive-fluence centroid provides the preferred timing measure because it uses the full positive-excess light curve and is less dependent on a single sampled maximum. For the combined AllWISE 2010 plus pre-rise NEOWISE baseline, the Monte Carlo resampling with bootstrap resampling of the baseline candidates gives $\Delta t_{\rm cent}=224\pm45$ d, corresponding to $c\Delta t_{\rm cent}=0.188\pm0.038$ pc (Table~\ref{tab:mir_timing_summary}). We adopt this value as the effective differential response scale
between the $W1$- and $W2$-weighted MIR emission. The difference between the Monte Carlo estimate and the corresponding fixed-baseline value reflects the bootstrap resampling of the pre-flare baseline candidates included in the uncertainty analysis.

The $W2$--$W1$ centroid offset persists across alternative baseline choices. In fixed-baseline recalculations, using only the first two pre-rise NEOWISE epochs gives $\Delta t_{\rm cent}\simeq207$ d, corresponding to $c\Delta t_{\rm cent}\simeq0.173$ pc; using the combined AllWISE 2010 plus NEOWISE epochs 0--1 baseline gives $\Delta t_{\rm cent}\simeq234$ d, corresponding to $c\Delta t_{\rm cent}\simeq0.196$ pc; and using only the two 2010 AllWISE epochs gives $\Delta t_{\rm cent}\simeq257$ d, corresponding to $c\Delta t_{\rm cent}\simeq0.215$ pc (Appendix~\ref{app:baseline_sensitivity}). Although the exact value depends slightly on the adopted baseline, all tested definitions consistently indicate a delayed $W2$ response on sub-parsec scales.

The $W2$ excess is also broader in time than $W1$.
Using the positive-excess flux-weighted RMS width defined in
Eq.~\ref{eq:positive_width}, we obtain approximately $697$ d for $W1$
and $781$ d for $W2$, corresponding to a width ratio of $\simeq1.12$.
The broader temporal width and the larger positive fluence describe
two distinct properties of the $W2$ response: its temporal extent and its
time-integrated excess.

We used the shift-and-broaden fit to test the consistency of a delayed, temporally extended dust-response scenario with the relative $W1$--$W2$ evolution. The fit gives a template shift of $\tau\simeq110\pm20$ d, corresponding to $c\tau\simeq0.092\pm0.017$ pc. This model-dependent shift characterizes the alignment of the light curves, rather than directly measuring the physical delay.

\subsection{Thermal evolution and IR energetics}
\label{subsec:teff_results}
\begin{figure}
\centering
\includegraphics[width=\columnwidth]{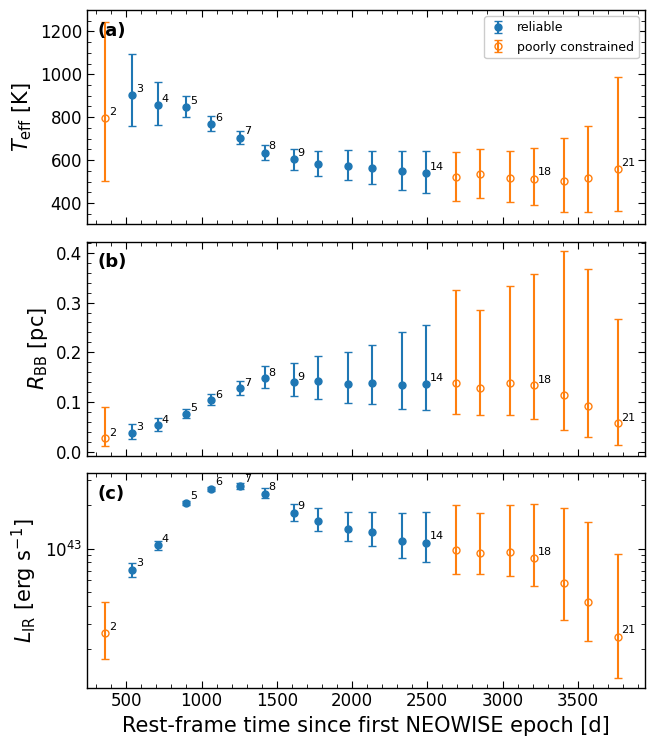}
\caption{
Evolution of the effective $W1$--$W2$ colour-blackbody properties of UGC~11487. From top to bottom: effective colour temperature, effective blackbody radius, and approximate bolometric IR luminosity. Filled symbols mark reliable epochs, while open symbols mark finite but non-reliable estimates. These quantities are derived from the two-band WISE excess fluxes and are used as effective diagnostics of the temperature, emitting scale, and luminosity evolution of the MIR flare.
}
\label{fig:wise_colour_blackbody}
\end{figure}

The effective $W1$--$W2$ colour-blackbody analysis provides a complementary empirical view of the thermal evolution of the MIR flare. Figure~\ref{fig:wise_colour_blackbody} shows the evolution of the effective colour temperature, blackbody radius, and approximate bolometric IR luminosity derived from the baseline-subtracted WISE excess fluxes.

Finite $T_{\rm eff}$ solutions are obtained for 20 non-baseline epochs. Of these, 12 epochs satisfy the adopted reliability cuts on the excess-flux signal-to-noise ratio and on the fractional uncertainties in $T_{\rm eff}$, $R_{\rm BB}$, and $L_{\rm IR}$. The remaining eight finite estimates are retained as poorly constrained, mainly because the $W1$ excess approaches the baseline at late times while $W2$ remains detectable. The first two NEOWISE epochs are used as part of the baseline definition and are therefore not assigned reliable colour-blackbody quantities.

The effective colour temperature decreases during the event. During the rise, $T_{\rm eff}$ is of order $900$ K. Near the MIR maximum it is typically $\sim700$--$850$ K, and during the decline it decreases to $\sim500$--$600$ K. Over the same period, the effective blackbody radius increases from $\sim0.03$--$0.05$ pc on the rise to $\sim0.1$--$0.15$ pc near and after the MIR maximum (Fig.~\ref{fig:wise_colour_blackbody}). This evolution suggests a transition from a hotter compact MIR component to a cooler and more persistent $W2$-dominated response. At late times, the uncertainties in $R_{\rm BB}$ and $L_{\rm IR}$ increase because the $W1$ excess approaches the baseline while $W2$ remains detectable. These late-time estimates are therefore retained as poorly constrained effective emitting-radius estimates and are not used to infer precise dust radii.

The approximate bolometric IR luminosity reaches $\sim3\times10^{43}$ erg s$^{-1}$. Integrating the effective $L_{\rm IR}(t)$ over rest-frame time gives $E_{\rm IR}\simeq2.9\times10^{51}$ erg for the fiducial reliable-epoch estimate. Including all finite non-baseline estimates gives $E_{\rm IR}\simeq3.8\times10^{51}$ erg. We therefore regard the two-band effective IR energy as being of order a few $10^{51}$ erg.

\rev{The decreasing effective temperature, increasing emitting scale, and persistent $W2$ emission are complementary indications of a thermally stratified, extended MIR response.}

\subsection{Phenomenological and physical descriptions of the MIR response}
\label{subsec:model_results}
\label{subsec:hydro_clumpy_results}

The smooth joint $W1+W2$ phenomenological modeling shows that
the observed WISE morphology cannot be adequately reproduced
by \rev{a single fixed-colour, self-similar component.} With the effective-error treatment,
the one-component model gives $\chi^2_{\rm eff}=321.70$ and is
strongly disfavored relative to the two-component model, with
$\Delta{\rm BIC}=281.7$ (Table~\ref{tab:model_comparison}).
The smooth two-component model provides the preferred compact
phenomenological description, with $\chi^2_{\rm eff}=21.06$,
${\rm BIC}_{\rm eff}=58.91$, and $\Delta{\rm BIC}=0$.

The three-component model reduces $\chi^2_{\rm eff}$ to 14.80
and follows the marginal late $W2$ shoulder more closely.
However, after penalizing the additional free parameters,
it is disfavored relative to the two-component model by
$\Delta{\rm BIC}=12.7$. We therefore adopt the smooth
two-component model as the preferred phenomenological
description of the WISE morphology, while treating the
three-component case as a diagnostic of possible late-time
residual structure. The Monte Carlo parameter estimates
for the preferred smooth two-component model are listed
in Table~\ref{tab:two_component_parameters}.

The two phenomenological components describe an early,
hotter contribution and a cooler, longer-lived contribution.
This decomposition motivates a spatially and thermally
distributed dust response. It does not uniquely determine
the number or boundaries of physical dust regions.

The staged search sampled all 141 fallback histories
in the adopted STARS input set
(Appendix~\ref{app:hydro_clumpy_details}).
We retain two candidate solutions, evaluated with the same
radiative-transfer solver and numerical resolution.
Case~A is the independently refined reference candidate,
while Case~B was selected from the staged grid search and
subsequently refined locally.
Their parameters and derived quantities are listed in
Table~\ref{tab:hydro_clumpy_models}.

Case~A gives $\chi^2=58.83$ for the nominal
conditional $\nu=37$, compared with $\chi^2=58.92$
for $\nu=37$ in Case~B. Their difference,
$\Delta\chi^2=0.09$, is smaller than the numerical
sensitivity found in the radial-resolution test.
The two solutions therefore have indistinguishable
fit quality at the precision of the present
calculation. We use Case~A as the reference
solution and retain Case~B as an alternative
physical configuration. The illuminating fallback
history remains degenerate with accretion smoothing
and dust-response parameters.

\begin{table*}
\centering
\small
\caption{Parameters and derived quantities of the
two refined physical dust-echo solutions.}
\label{tab:hydro_clumpy_models}
\begingroup
\setlength{\tabcolsep}{7pt}
\renewcommand{\arraystretch}{1.40} 
\begin{tabular}{lcc}
\hline
Quantity & Case A & Case B \\
\hline
\multicolumn{3}{c}{Selected fallback history} \\
\hline
STARS identifier
& \texttt{m3.0\_t0.0/0.850.dat}
& \texttt{m0.7\_t1.0/0.800.dat} \\
\hline
\multicolumn{3}{c}{Fitted continuous parameters} \\
\hline
Radiative efficiency $\eta$
&  $0.15^{+0.01}_{-0.01}$ \\

Accretion-smoothing time $\tau_{\rm visc}$ (d)
&  $240.87^{+31.93}_{-33.19}$
&  $245.25^{+29.38}_{-36.83}$ \\

Compact-population optical depth $\tau_{\rm C}$
&  $0.38^{+0.03}_{-0.03}$
&  $0.39^{+0.03}_{-0.03}$ \\

Extended-population optical depth $\tau_{\rm E}$
&  $1.91^{+0.71}_{-0.51}$
&  $1.98^{+0.60}_{-0.60}$ \\

Polar opening angle $\theta_{\rm open}$ (deg)
&  $46.79^{+2.47}_{-3.47}$
&  $46.91^{+2.34}_{-3.54}$ \\

Inclination $i$ (deg)
&  $42.61^{+2.37}_{-2.52}$
&  $42.66^{+2.32}_{-2.67}$ \\

Event epoch $t_0$ (MJD)
& 57067.08
& 57065.25 \\
\hline
\multicolumn{3}{c}{Fixed properties shared by both solutions} \\
\hline
$\log(M_{\rm BH}/M_\odot)$
& \multicolumn{2}{c}{6.8} \\
Compact $(R_{\rm in},R_{\rm out})$ (pc)
& \multicolumn{2}{c}{$(0.25,0.50)$} \\
Extended $(R_{\rm in},R_{\rm out})$ (pc)
& \multicolumn{2}{c}{$(1.00,2.00)$} \\
Radial density indices $(p_{\rm C},p_{\rm E})$
& \multicolumn{2}{c}{$(0,2)$} \\
Grain composition
& \multicolumn{2}{c}{Graphite} \\
Grain-size interval ($\mu$m)
& \multicolumn{2}{c}{$0.1$--$1.0$} \\
Grain-size index $q$
& \multicolumn{2}{c}{3.5} \\
Fixed $W1$ host flux density (mJy)
& \multicolumn{2}{c}{8.072} \\
Fixed $W2$ host flux density (mJy)
& \multicolumn{2}{c}{5.215} \\
\hline
\multicolumn{3}{c}{Fit statistics and derived quantities} \\
\hline
Joint $\chi^2$
& 58.83 & 58.92 \\

Nominal conditional $\nu$ & 37 & 37 \\

$W1$ contribution to $\chi^2$
& 19.11 & 19.26 \\

$W2$ contribution to $\chi^2$
& 39.72 & 39.67 \\

Maximum grain temperature (K)
& 1142.80 & 1139.76 \\

$E_{\rm src,win}$ ($10^{51}$ erg)
& 8.49 & 8.54 \\

$E_{\rm src,abs}$ ($10^{51}$ erg)
& 5.19 & 5.25 \\

$E_{\rm IR,esc}$ ($10^{51}$ erg)
& 5.19 & 5.25\\

$E_{\rm src,esc}$ ($10^{51}$ erg)
& 3.30 & 3.29 \\

$E_{\rm src,abs}/E_{\rm src,win}$
& 0.61 & 0.61 \\
\hline
\end{tabular}
\endgroup
\tablefoot{Central values are the adopted direct-RT estimates. The six asymmetric ranges are approximate conditional one-parameter variations near $\Delta\chi^2=1$, obtained with response-emulator profiles and selected direct endpoint checks. The energy quantities refer to the computed source window and the eventual escape associated with the computed
emission; escaping energies are integrated over all directions. The maximum grain temperatures and energy quantities
are evaluated at the adopted parameter values; their
uncertainties have not been propagated.
}
\end{table*}

\begin{figure}
\centering
\includegraphics[width=\columnwidth]
{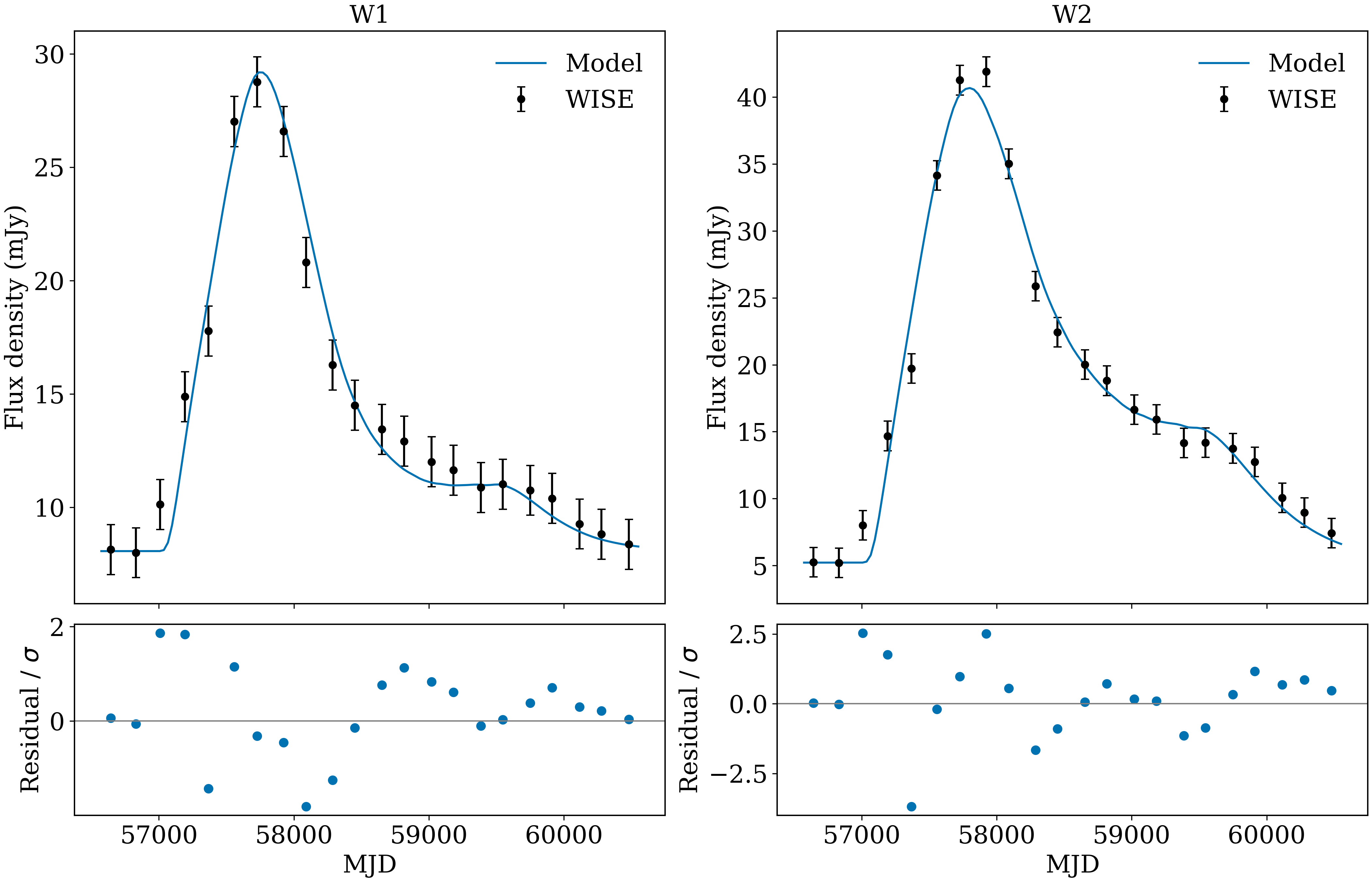}
\caption{
Joint radiative-transfer fit to the WISE/NEOWISE light curves
for \rev{the refined Case~A solution in Table~\ref{tab:hydro_clumpy_models}.} The upper panels show the observed total flux
densities and the model, including the fixed host contribution,
in $W1$ and $W2$. Error bars show the effective photometric uncertainties used in the fit.
}
\label{fig:rt_case_a_lightcurves}
\end{figure}

The refined Case~A light curves reproduce the broad rise,
maximum, and decade-long decline in both WISE bands
(Fig.~\ref{fig:rt_case_a_lightcurves}).
The residuals retain phase-dependent structure, with a larger
contribution from $W2$.
Using the convention
$r=(F_{\rm obs}-F_{\rm model})/\sigma_{\rm eff}$,
the largest absolute residual is \rev{$r\simeq-3.69$} in $W2$
at MJD~57367.67, where the model over predicts the rising
light curve. Positive $W2$ residuals of approximately
$2.53\sigma$ and \rev{$2.50\sigma$} occur at MJD~57008.22
and 57923.00, respectively.
The remaining mismatch therefore includes the onset and
peak region as well as the subsequent decline.

At the adopted parameter values, the maximum grain
temperatures are 1142.80~K in Case~A and 1139.76~K in Case~B.
Both solutions remain below the adopted sublimation threshold
of approximately 1793~K and satisfy the solver's
static-opacity criterion.
These maxima characterize the hottest grains in the computed
temperature field. The empirical two-band colour temperature
instead summarizes the integrated excess-flux ratio under
the adopted blackbody approximation.
The geometric dust radii, effective blackbody emitting radius,
and differential interband response scales consequently
describe distinct properties of the system.

\begin{figure}
\centering
\includegraphics[width=\columnwidth]
{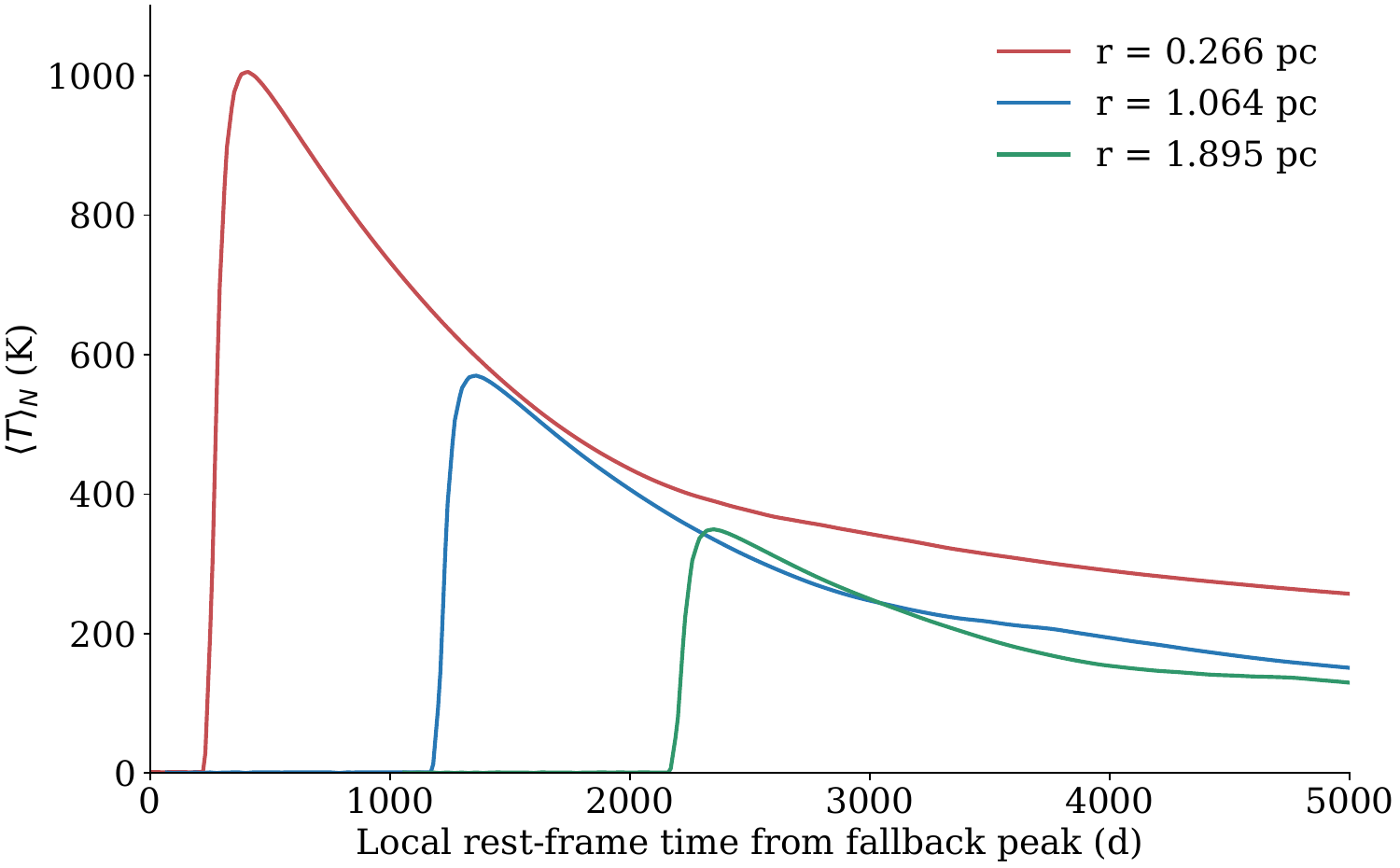}
\caption{
Local thermal evolution in \rev{the refined Case~A solution} at representative radii
of 0.266, 1.064, and 1.895~pc.}
\label{fig:case_a_temperature_time}
\end{figure}

\begin{figure*}
\centering
\includegraphics[width=\textwidth]
{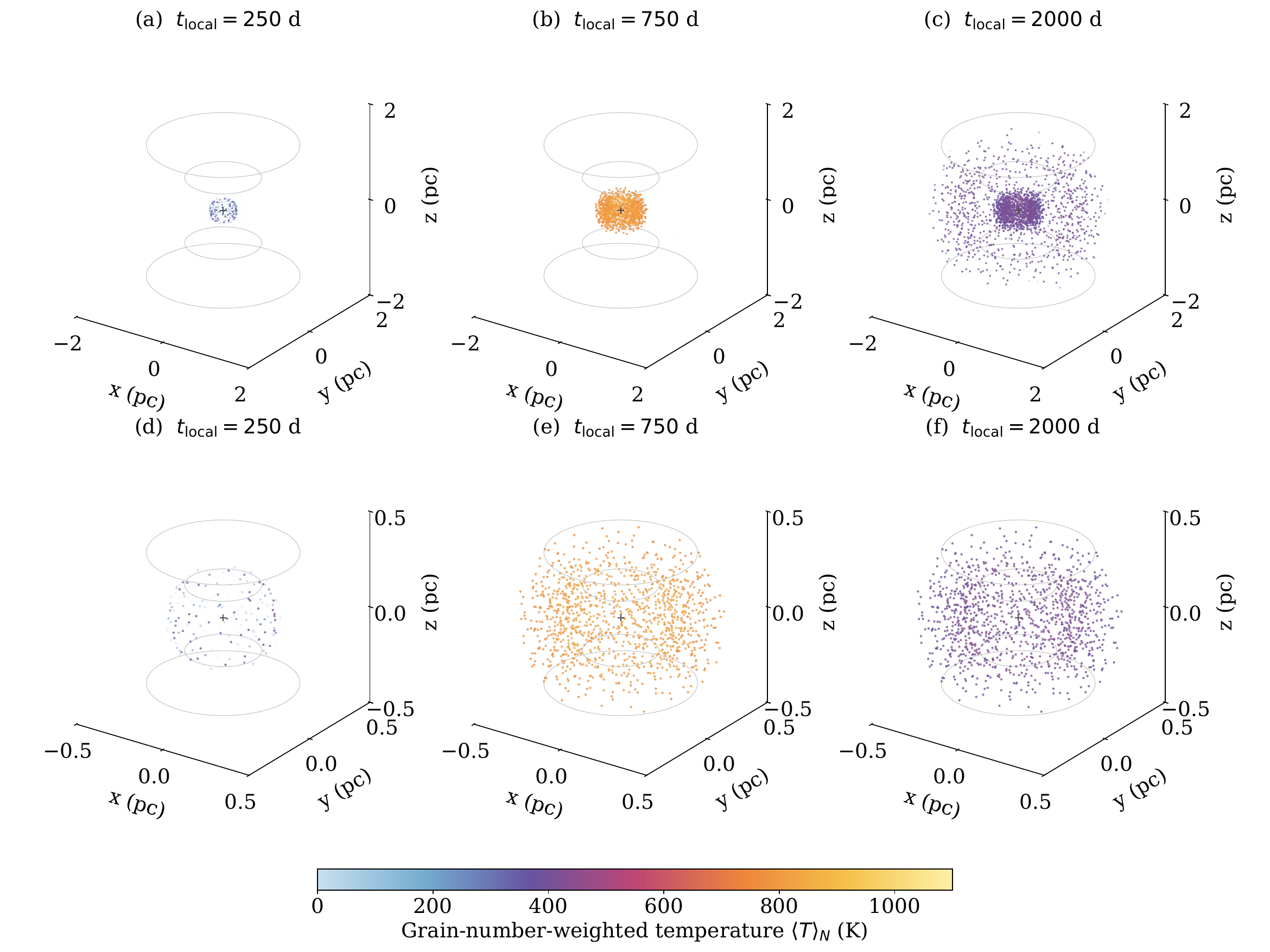}
\caption{
Three-dimensional representation of the \rev{Case~A} temperature
field at simultaneous local rest-frame times of
250, 750, and 2000~d.
\rev{The upper row shows the full distribution; the lower row enlarges the compact population.}\rev{The central marker indicates the illuminating source.} Points sample the effective smooth dust distribution;
their plotted density does not encode the physical
grain-number density or identify individual dust clouds.
These snapshots show the local thermal state, before
integration over observer-dependent arrival times.
}
\label{fig:case_a_thermal_3d}
\end{figure*}

The thermal evolution of the refined Case~A solution is illustrated in Figs.~\ref{fig:case_a_temperature_time} and~\ref{fig:case_a_thermal_3d}.
The temperature varies continuously within each radial population and depends on grain size. The displayed thermal snapshots use a common local rest-frame time, $t_{\rm loc}$, measured relative to the peak of the
unsmoothed fallback history. For a location at radius $r$, the saved source-retarded coordinate is $u=t_{\rm loc}-r/c$. The observed light curves additionally include the position-dependent propagation time to the observer and cosmological time dilation.

Figure~\ref{fig:case_a_temperature_time} shows the
grain-number-weighted temperature at three representative
radial cells. At $r=0.266$~pc, the mean temperature reaches
approximately \rev{1005~K} at $t_{\rm loc}\simeq410$~d.
At $r=1.064$ and 1.895~pc, the corresponding maxima are
approximately \rev{570 and 349~K}, reached at
$t_{\rm loc}\simeq1360$ and 2350~d.
These values describe the computed local temperature
histories; the quoted peak times are read from the sampled
curves. The later heating at larger radii reflects the
propagation of the illuminating radiation through the
extended dust distribution.

The simultaneous three-dimensional snapshots
(Fig.~\ref{fig:case_a_thermal_3d}) place this thermal evolution
within the adopted axisymmetric geometry.
The dust distribution remains spatially fixed, while the
temperature field changes as the radiation propagates
outwards and the illuminating source evolves.
The moving pattern of heating therefore represents a
radiative response, not outward motion of the dust.

The refined solutions have the source and escaping energies listed in Table~\ref{tab:hydro_clumpy_models}. Their relative global energy-budget residuals are $7.28\times10^{-7}$ (A) and $9.14\times10^{-7}$ (B), demonstrating internal energy closure for these calculations, not uniqueness or resolution convergence.

The corresponding absorbed fractions are approximately
0.61 in both cases. These fractions depend on the angular
coverage and finite absorption optical depths of the dust.
The escaping model energies are integrated over the computed
dust spectrum and all directions, including the response
beyond the observed epochs.

\subsection{Possible late-time \texorpdfstring{$W2$}{W2} excess and MIR--radio scale comparison}
\label{subsec:late_bump_results}
\label{subsec:radio_results}

We tested whether the apparent late-time $W2$ residual around epochs 17--19 in Fig.~\ref{fig:wise_lightcurve} represents a significant secondary excess. We fitted the surrounding post-peak $W2$ decline with a smooth exponential-plus-constant profile after excluding epochs 17--19, and then measured the residuals of the excluded epochs relative to this smooth decline. The details of this local residual test are given in Appendix~\ref{app:baseline_sensitivity}.

For the fiducial exponential-plus-constant decline, the
weighted $W2$ residual over epochs 17--19 is
$1.49\pm0.72~{\rm mJy}$, corresponding to a formal significance of
$2.06\sigma$. The one-sided parametric-bootstrap probability is
$p=0.054$. The corresponding $W1$ residual is
$0.59\pm0.65~{\rm mJy}$, or $0.92\sigma$. The largest positive
$W2$ residuals occur at epochs 17 and 18, while epoch 19 is consistent
with the fitted decline. We therefore identify this interval as a
possible late-time $W2$ excess with modest statistical significance.
For the MIR--radio travel-time comparison, its midpoint is defined as
the median MJD of epochs 17--19,
${\rm MJD}=59911.004$, corresponding to epoch 18.

\begin{figure}
\centering
\includegraphics[width=\columnwidth]{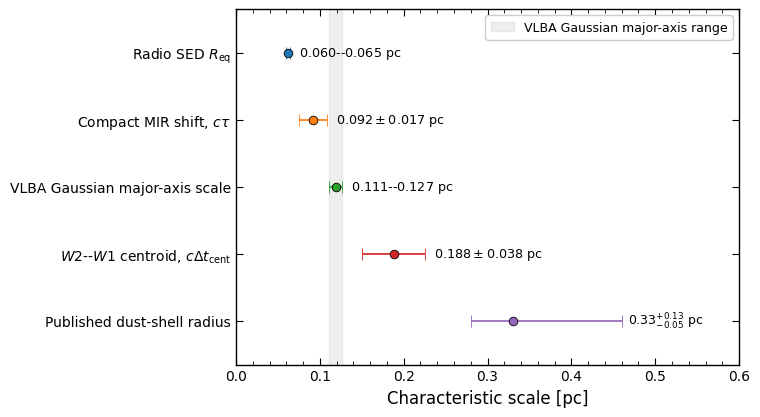}
\caption{
Comparison of characteristic MIR and radio scales.
The radio SED range represents the cooling-break-corrected
equipartition radii reported by \citet{Golay2026}, while the shaded
region marks the VLBA Gaussian major-axis range of
$0.111$--$0.127$ pc. The $W2$--$W1$ centroid scale represents an
effective differential response scale between the $W1$- and
$W2$-weighted MIR emission, while the shift-and-broaden
$W1$-to-$W2$ scale gives the shorter compact-shift diagnostic.
The figure places the thermal MIR dust-response diagnostics and the
synchrotron radio-emitting region within the central sub-parsec nuclear
environment.
}
\label{fig:mir_radio_scales}
\end{figure}

We also compared the MIR-derived scales with the radio properties reported by
\citet{Golay2026}, which independently probe the compact nuclear environment
of WTP14adeqka.
The published radio constraints are summarized in
Table~\ref{tab:target_properties}. The cooling-break-corrected radio SED
equipartition radii are $R_{\rm eq}=0.065\pm0.003$ and
$0.060\pm0.003$ pc for the two multifrequency epochs, while the VLBA
Gaussian major-axis scale increases from $0.111$ to $0.127$ pc.
The preferred model-independent $W2$--$W1$ centroid scale,
$c\Delta t_{\rm cent}=0.188\pm0.038$ pc,
represents an effective differential response scale between the
$W1$- and $W2$-weighted MIR emission,
whereas the shift-and-broaden $W1$-to-$W2$ analysis gives a shorter
compact-shift scale, $c\tau=0.092\pm0.017$ pc.
The compact-shift scale lies between the radio SED equipartition
radii and the VLBA Gaussian major-axis scale. This scale ordering supports a structured sub-parsec nuclear environment in which WISE traces thermal dust reprocessing over a radially or thermally extended dust distribution, while the radio emission traces synchrotron-emitting outflow or shock material on comparable compact nuclear scales.
The compact-shift, centroid, and literature-shell scales
characterize complementary MIR scales used in the comparison.
Table~\ref{tab:radio_travel_time} gives the corresponding required
travel-time velocities, $\beta_{\rm req}$, adopting 2015 April 14
(${\rm MJD}=57126$) as the common reference epoch and converting the
elapsed intervals to the source rest frame.
The relative placement of the characteristic MIR and radio scales is shown in Fig.~\ref{fig:mir_radio_scales}. Their possible physical connection is discussed in Sect.~\ref{subsec:discussion_radio_energy}.

\subsection{Time-averaged host-galaxy properties from UV-to-radio SED modeling}
\label{subsec:sed_results}
\begin{figure}
\centering
\includegraphics[width=\columnwidth]{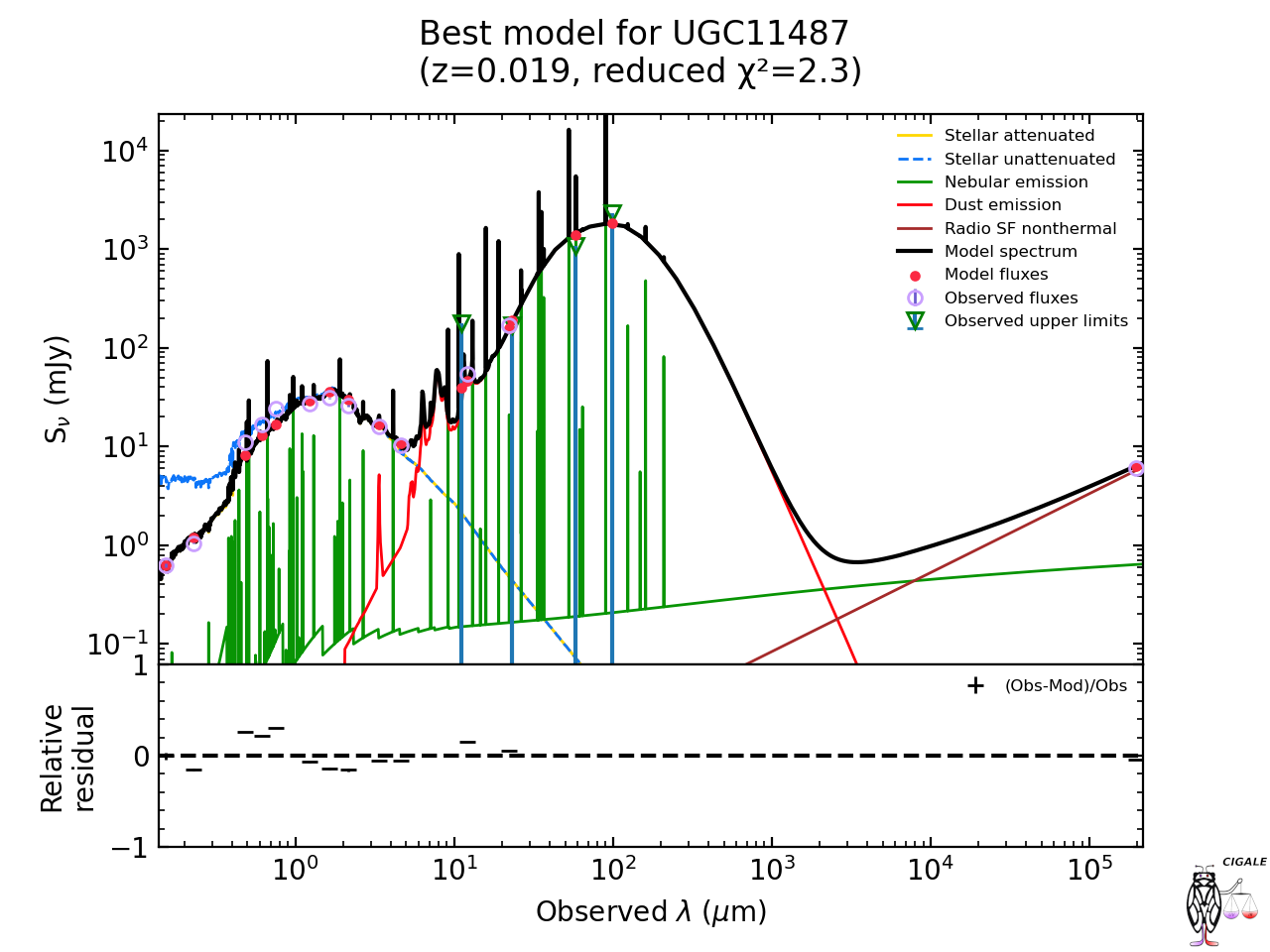}
\caption{
Best-fitting CIGALE model for the broadband UV--radio SED of UGC~11487. The fit is based on the aperture-matched host-galaxy photometry described in Sect.~\ref{subsec:sed_photometry_methods}. All IRAS measurements at 12, 25, 60, and 100 $\mu{\rm m}$ are upper limits.
}
\label{fig:ugc11487_sed_fit}
\end{figure}

The broadband UV--radio SED of UGC~11487 was modeled with CIGALE using the aperture-matched photometry described in Sect.~\ref{subsec:sed_photometry_methods} and in detail in Appendix~\ref{app:host_sed_methods}. The initial grid parameters and the CIGALE modules used for SED modeling are described in Appendix~\ref{app:CIGALE_input}. The fit reproduces the observed host-galaxy SED with a reduced $\chi^2_\nu\simeq2.27$ (Fig.~\ref{fig:ugc11487_sed_fit}; Table~\ref{tab:cigale_host_properties}). The Bayesian estimates indicate a moderately star-forming host galaxy, with $M_\star=(2.01\pm0.15)\times10^{10}\,M_\odot$, ${\rm SFR}=2.92\pm0.15\,M_\odot\,{\rm yr}^{-1}$, and ${\rm sSFR}\simeq1.45\times10^{-10}\,{\rm yr}^{-1}$. 

The SED fit gives $E(B-V)_\star=0.176\pm0.001$ and $E(B-V)_{\rm lines}=0.400\pm0.003$. The corresponding FUV attenuation is $A_{\rm FUV}=2.17\pm0.02$ mag. The infrared emission is dominated by star-formation-heated dust, with $L_{\rm dust}=(1.05\pm0.05)\times10^{37}\,{\rm W}$, $\log(L_{\rm dust}/L_\odot)\simeq10.44$, and $\alpha_{\rm dust}=1.67\pm0.08$, placing UGC~11487 below the classical LIRG regime. \rev{All four IRAS measurements, at 12, 25, 60, and 100 $\mu{\rm m}$, are treated as upper limits, as described in Appendix~\ref{app:host_sed_methods}.}

Applying the local AGN black-hole--total-stellar-mass
relation of \citet{ReinesVolonteri2015} to our SED-derived
stellar mass gives $\log_{10}(M_{\rm BH}/M_\odot)\simeq6.72$.
This differs by $0.08$~dex from the adopted value of
$6.8\pm0.5$ from \citet{Masterson2024}, well within the
$0.55$~dex empirical rms scatter of the relation.
Both estimates use the same scaling relation; the comparison
therefore provides a consistency check based on our separate
host-SED analysis.

Overall, the host-galaxy properties are consistent with a moderately star-forming isolated barred disc in which the MIR flare appears as a transient phenomenon superposed on an otherwise normal galaxy.

\begin{table}
\centering
\caption{CIGALE-derived host-galaxy properties of UGC~11487.}
\label{tab:cigale_host_properties}
\setlength{\tabcolsep}{3.5pt}
\begin{tabular}{l c}
\hline
Quantity & Value \\
\hline
Reduced $\chi^2_\nu$ & $\simeq2.27$ \\
$M_\star$ & $(2.01\pm0.15)\times10^{10}\,M_\odot$ \\
SFR & $2.92\pm0.15\,M_\odot\,{\rm yr}^{-1}$ \\
sSFR & $\simeq1.45\times10^{-10}\,{\rm yr}^{-1}$ \\
$E(B-V)_\star$ & $0.176\pm0.001$ \\
$E(B-V)_{\rm lines}$ & $0.400\pm0.003$ \\
$A_{\rm FUV}$ & $2.17\pm0.02$ mag \\
$L_{\rm dust}$ & $(1.05\pm0.05)\times10^{37}\,{\rm W}$ \\
$L_\star$ & $(2.053\pm0.103)\times10^{37}\,{\rm W}$ \\
$\log(L_{\rm dust}/L_\odot)$ & $\simeq10.44$ \\
$\alpha_{\rm dust}$ & $1.67\pm0.08$ \\
\hline
\end{tabular}
\end{table}

\section{Discussion}
\label{sec:discussion}
\subsection{UGC~11487 as an obscured TDE-like nuclear flare}
\label{subsec:discussion_mir_context}

UGC~11487/WTP14adeqka is a long-lived MIR-selected nuclear flare whose evolution is dominated by dust-reprocessed emission. The decade-scale WISE light curve shows a larger peak-to-peak amplitude and a larger time-integrated excess in $W2$ than in $W1$, together with an RWB trend and phase-dependent flux--flux behaviour (Table~\ref{tab:mir_summary}). These properties are consistent with a transient heating event followed by a structured dust response, as expected when a substantial fraction of the primary UV/optical radiation from a nuclear transient is absorbed and re-emitted by circumnuclear dust \citep{Jiang2016, Jiang2021, Masterson2024, Necker2025}.

The MIR evolution is consistent with a dust-obscured nuclear transient and, in particular, with a TDE-like accretion episode reprocessed by circumnuclear dust. A definitive classification as a classical TDE is limited by the absence of a directly and densely sampled primary optical/UV flare. Alternative nuclear-accretion scenarios, including an accretion-state change or a changing-look AGN episode, therefore remain viable. We consequently interpret UGC~11487 as an obscured TDE candidate or TDE-like nuclear flare.

A supernova embedded in a dusty circumstellar environment can
also produce strong infrared emission through an IR echo, newly formed
dust, or continued circumstellar interaction. However, the energetics
and duration of UGC~11487 are extreme for this interpretation. The effective bolometric IR energy inferred from the two-band MIR photometry is $E_{\rm IR}\simeq(2.9$--$3.8)\times10^{51}$ erg. Since this
represents reprocessed emission, the intrinsic radiated energy would be
larger for a dust covering factor below unity. Mid-infrared studies of
superluminous supernovae typically find dust emission evolving on
timescales of several hundred to approximately one thousand days
\citep{Sun2022SLSN}. Rare, exceptionally energetic interacting
supernovae do exist: SN~2006gy remained luminous in the infrared for
several years \citep{Fox2015SN2006gy}, while SN~2016aps radiated
$\gtrsim5\times10^{51}$ erg in total and required an explosion energy
above $10^{52}$ erg \citep{Nicholl2020SN2016aps}. The energy budget
therefore does not exclude every possible extreme supernova, but it
strongly disfavors an ordinary dusty SN. Together with the nuclear
location, the decade-long MIR evolution, and the delayed resolved
non-relativistic radio outflow \citep{Golay2026}, the observed
properties favor a TDE-like nuclear accretion event over a supernova
origin.

This interpretation is consistent with the growing sample of
obscured MIR-selected TDE candidates identified through their infrared
dust echoes \citep{Jiang2021MIRONG,Liu2026}. In particular,
\citet{Liu2026} showed that detailed infrared observations can recover
the properties of an otherwise hidden nuclear flare in a star-forming
galaxy, further demonstrating the diagnostic power of MIR dust echoes
for obscured nuclear transients.

\subsection{Host-galaxy context and the transient nature of the MIR flare}
\label{sec:host_context_discussion}

The host-galaxy SED modeling presented in Sect.~\ref{subsec:sed_results} provides an important context for interpreting the MIR flare. The time-averaged UV--far-infrared emission of UGC~11487 is consistent with a moderately star-forming barred disc whose broadband energy budget is dominated by stellar light and star-formation-heated dust. An absent or weak persistent AGN contribution is an adopted model constraint, not an independent inference from this SED fit.

The host model provides a reference for the transient MIR component. The decade-long WISE flare, strong $W2$ excess, colour hysteresis, and delayed $W2$ response cannot be explained as static features of the host-galaxy SED and require an additional time-dependent nuclear component. UGC~11487 therefore appears to be a moderately star-forming isolated barred disc that underwent a temporary nuclear dust-reprocessing episode, compatible with an obscured TDE-like or other short-lived nuclear accretion event.

\subsection{TDEs as transient activity episodes in Milky Way analogue galaxies}
\label{sec:MWA}

TDEs in the Milky Way would be associated with Sgr~A*, although no Galactic TDE has been directly observed to date. Evidence for episodic past activity of Sgr~A* has been obtained from X-ray observations of molecular clouds in the Galactic Center region. In particular, studies by Sunyaev and collaborators demonstrated that several dense clouds, including Sgr~B2, act as natural X-ray reflection nebulae, reprocessing radiation emitted during past high-luminosity flares of the Galactic SMBH \citep{Revnivtsev2004, Churazov2017, Khabibullin2022, Marin2023}. Variability, spectral properties, and polarization measurements of the reflected emission indicate that Sgr~A* experienced significantly higher activity in the recent past than in its present quiescent state.

In general, the expected TDE rate is low, of the order of $\Gamma_{\rm TDE}\sim10^{-5}-10^{-4}\,{\rm yr}^{-1}$, corresponding to approximately one event per $10^{4}$--$10^{5}$ yr \citep{Komossa2015, Alexander2005}. Nevertheless, episodic TDE-driven energy injection has been discussed as a possible contributor to the past activity of Sgr~A* and to large-scale Galactic-center structures such as the Fermi and eROSITA bubbles \citep{Sazonov2012, BlandHawthorn2013, Su2010, Predehl2020}.

Recent work suggests that large-scale host structure may be
relevant to the enhanced occurrence of TDEs in some galaxy populations.
\citet{Chang2026TDEHosts} found that TDE hosts are more centrally
concentrated than matched control galaxies and are approximately
$1.5$--$2.5$ times more likely to show bar-like or ring-like structures,
with an even stronger enhancement among green-valley hosts. They
therefore proposed bar-driven secular evolution as a possible mechanism
for building the central stellar concentrations associated with enhanced
TDE rates. UGC~11487 is particularly relevant in this context: our
structural analysis securely detects a large-scale stellar bar with
$R_{\rm bar}=4.76^{+0.53}_{-0.34}$ kpc
($p_{\rm bar}\simeq2\times10^{-4}$), together with a projected central
stellar surface density of
$\log\Sigma_{1{\rm kpc}}\simeq9.03$--$9.19$
($M_\odot\,{\rm kpc^{-2}}$). These properties are consistent with
secular bar-driven evolution contributing to the build-up of the central
stellar distribution. However, a direct causal link between a
large-scale bar and the occurrence of an individual TDE remains
uncertain; \citet{Nicholl2020AT2019qiz} similarly noted the prominent
bar in the host of AT~2019qiz while cautioning that large-scale bars
have not been demonstrated to enhance the TDE rate.

In this context, TDEs may represent short-lived episodes in the evolution of otherwise normal spiral galaxies. Systems such as UGC~11487 are therefore of particular interest, because they are candidates for energetic accretion episodes in otherwise quiescent Milky-Way-like galaxies \citep{Vavilova2024}. Such episodes may provide a glimpse of processes that could have occurred in the recent history of the Milky Way itself. Consequently, TDE candidates among isolated barred galaxies offer a useful bridge between studies of transient accretion phenomena and evolutionary diagnostics of MWNeTs \citep{Vavilova2026, Vavilova2026b,Dmytrenko2025,Fedorov2026}

\subsection{Structured dust response and TDE-like reprocessing}
\label{subsec:discussion_structured_dust}

The empirical diagnostics, phenomenological decomposition, and radiative-transfer solutions provide complementary descriptions of the same $W1$ and $W2$ observations. Their comparison tests whether the physical model provides a consistent explanation of the observed thermal and temporal evolution. Since these methods use the same photometry, this is an internal consistency check, not an independent validation.

The phenomenological decomposition identifies an earlier, hotter contribution and a cooler, longer-lived contribution. This behaviour is qualitatively consistent with the Case~A temperature histories, which show earlier, stronger heating of the compact population and a later, cooler response at larger radii (Figs.~\ref{fig:case_a_temperature_time} and~\ref{fig:case_a_thermal_3d}). Each radial population contains grains with different temperatures and contributes to both WISE bands. The decreasing empirical colour temperature describes their integrated two-band emission and need not match the hottest grains in the physical model. Changes in the effective blackbody radius can likewise reflect changing contributions from different parts of a spatially fixed dust distribution.

The timing diagnostics and model temperature histories describe different aspects of this evolution. Local heating maxima include the propagation time of the illuminating radiation, while the observed light curves also include travel times from the emitting dust to the observer. The centroid offset and shift-and-broaden lag measure differences between the wavelength-weighted responses in the two bands. Because these bands receive emission from overlapping regions, the inferred differential scales can be smaller than the dust radii within the adopted compact and extended populations, 0.25--0.50 and 1.00--2.00~pc, respectively. They do not directly measure the separation between these populations. The preferred two phenomenological components therefore do not uniquely establish two sharply bounded dust structures or an empty radial gap.

The refined Case~A solution reproduces the broad rise, maximum, and decade-long decline in both bands, while retaining phase-dependent residuals, particularly in $W2$ (Fig.~\ref{fig:rt_case_a_lightcurves}). Together with the qualitative thermal comparison, this supports the model as a description of the broad dust-response evolution. The detailed local heating histories remain predictions conditional on the adopted source history and dust geometry. Cases~A and B illustrate this dependence: different fallback histories yield nearly identical fit statistics. Their effective accretion-smoothing times of approximately 240--245~d do not separately resolve circularization and inward transport or uniquely determine the disrupted stellar properties.

The total escaping dust-emission energies in both model solutions are comparable to the empirical bolometric infrared energy estimates. The absorbed source-energy fraction of approximately 0.61 combines angular coverage, finite optical depth, and the adopted spectrum and is not a purely geometric covering factor.

In both solutions, $i<\theta_{\rm open}$, so the modeled dust does not obscure the central source. Additional cooler dust along the line of sight could provide attenuation \citep{Masterson2024,Masterson2025JWST}, but is not separately modeled here. The aperture-matched photometry and host-galaxy SED analysis do not robustly constrain its mass or spatial distribution because the far-infrared constraints in our dataset consist only of IRAS upper limits. The integrated host attenuation also does not determine nuclear obscuration. The contribution of additional dust to the absence of a detected optical flare therefore remains to be tested.

\subsection{MIR--radio connection and energetic consistency}
\label{subsec:discussion_radio_energy}

Independent radio observations of WTP14adeqka provide an external probe of the compact nuclear environment \citep{Golay2026}. The published radio constraints summarized in Table~\ref{tab:target_properties} provide an independent comparison with the MIR-derived response scales and connect the thermal dust response with synchrotron-emitting plasma associated with an outflow or shock \citep{Alexander2020RadioTDEs,Roth2020Radiative}.

Delayed radio emission is increasingly recognized as an important stage of TDE evolution. Radio emission emerging hundreds to thousands of days after discovery has been reported in both optically and X-ray-selected TDEs, favoring delayed outflows in many systems \citep{Cendes2024LateRadio,Goodwin2025RadioXrayTDEs}. Long-lived radio emission may contain multiple temporal components and include host contamination \citep{Christy2026LongLivedRadio}. Years-delayed radio flares in slowly evolving TDE candidates provide a relevant comparison for WTP14adeqka \citep{Zhang2026YearsDelayedRadio}.

The cooling-break-corrected radio SED analysis gives equipartition radii of $R_{\rm eq}=0.060$--$0.065$ pc, and the VLBA elliptical-Gaussian fits give major-axis scales of $0.111$--$0.127$ pc \citep{Golay2026}. The compact MIR shift scale, $c\tau=0.092\pm0.017$ pc, lies between these characteristic radio scales. The $W2$--$W1$ centroid scale, $c\Delta t_{\rm cent}=0.188\pm0.038$ pc, traces a larger effective differential response scale. Their relative placement is shown in Fig.~\ref{fig:mir_radio_scales}.

The possible late-time $W2$ excess over epochs 17--19 has a midpoint at ${\rm MJD}=59911.00425$ and occurs during the radio-bright phase. Its weighted residual has a formal significance of $2.06\sigma$, with a one sided parametric-bootstrap probability of $p=0.054$. Table~\ref{tab:radio_travel_time} gives the required travel-time velocities from the common reference epoch, 2015 April 14 (${\rm MJD}=57126$), after conversion to the source rest frame. The compact MIR scale requires $\beta_{\rm req}\simeq0.040$ by the midpoint of the late-$W2$ interval, comparable to the published VLBA major-axis expansion velocity of $\beta\simeq0.049$. The corresponding velocities are $\beta_{\rm req}\simeq0.082$ for the centroid scale and $\simeq0.144$ for the literature-shell scale. The published radio SED and position-angle-averaged VLBA constraints are $\beta\simeq0.022$--$0.031$ and $\beta\simeq0.025$, respectively \citep{Golay2026}.

Several physical scenarios can account for delayed radio activity. In the debris--torus collision model of \citet{Lei2024}, unbound TDE debris impacts a dusty torus and produces a delayed radio outburst. \citet{Linial2026} proposed that viscous expansion of the TDE accretion disc can lead to repeated collisions with a pre-existing stellar EMRI, launching a slow, energetic outflow after an intrinsic delay. They identified WTP14adeqka as a plausible example of this mechanism. An additional possibility is delayed escape of an initially misaligned jet. In the model of \citet{TeboulMetzger2023}, rapid precession can prevent the jet from penetrating the surrounding disc-wind ejecta until hydrodynamic alignment of the disc--jet system with the SMBH spin, delaying jet escape and the associated radio flare by months to years. This framework was applied to AT~2018hyz. A similar process could contribute to the delayed radio onset of WTP14adeqka. Distinguishing this scenario from intrinsically delayed ejection or outflow--circumnuclear-medium interaction, and testing it against the compact, non-relativistic expansion inferred for WTP14adeqka, requires event-specific modeling \citep{Golay2026}. Interaction of promptly produced debris or an intrinsically
delayed outflow with compact circumnuclear material could contribute
to late-time dust heating.

The cooling-break-corrected radio-SED analysis gives $E_{\rm eq,radio}\simeq6.3\times10^{49}$~erg under the adopted microphysical assumptions \citep{Golay2026}. Powering the full effective bolometric IR output, $E_{\rm IR}\simeq(2.9$--$3.8)\times10^{51}$~erg, through shocks would require $E_{\rm kin}\gtrsim E_{\rm IR}/\epsilon_{\rm IR}$, where $\epsilon_{\rm IR}$ is the fraction of dissipated kinetic energy emitted in the infrared. Even for $\epsilon_{\rm IR}=1$, this is approximately 46--60 times the radio-inferred energy. The latter is not a model-independent upper limit on the total kinetic energy, and the first radio detection does not determine the outflow launch epoch. Outflow-related heating may contribute, but its magnitude and association with the tentative late-time $W2$ excess remain unestablished.

For $\eta_{\rm rad}=0.1$ and $f_{\rm cov}=0.1$--$0.5$, the effective IR energy implies $M_{\rm acc}\simeq0.032$--$0.21\,M_\odot$. Assuming $f_{\rm acc}=0.1$--$0.5$ gives an illustrative disrupted-star mass scale of
$M_{\star,\rm scale}\simeq0.065$--$2.1\,M_\odot$.

\subsection{Limitations and implications}
\label{subsec:discussion_limitations}

The main limitations follow directly from the available data. WISE provides only two MIR bands, so $T_{\rm eff}$, $R_{\rm BB}$, $L_{\rm IR}$, and $E_{\rm IR}$ are effective two-band colour-blackbody diagnostics of the MIR evolution. The NEOWISE cadence is sparse, making the positive-fluence centroid a more stable timing measure than the discrete $W1$ and $W2$ peak epochs. The pre-flare baseline also introduces a systematic uncertainty, which motivated the baseline-sensitivity tests of the $W2$--$W1$ centroid scale.

The physical inference is conditional on the adopted black-hole mass, fixed illuminating spectral shape, graphite optical properties, grain-size distribution, and axisymmetric density prescription. The uncertainty in $M_{\rm BH}$ is not propagated by holding it at $10^{6.8}\,M_\odot$. The calculations include absorption and thermal re-emission, while scattering, explicit clump substructure, and dynamical grain destruction are outside the implemented model.

The static-opacity temperature check identifies configurations for which the adopted dust population remains below its prescribed sublimation threshold. Configurations that fail this check require a calculation with evolving grain abundances and opacity. Their unsupported status does not constitute an observational exclusion.

Parameter intervals obtained within a selected fallback history and fixed radial geometry are approximate conditional variations, not statistically calibrated $1\sigma$ uncertainties. Differences between alternative fallback histories, baseline choices, and dust prescriptions contribute
additional model dependence. The finite parameter search and the two-band data do not uniquely determine the disrupted-star properties or the three-dimensional dust distribution. The fit evaluates the compatibility of TDE-driven source histories with the MIR observations. It does not provide a likelihood comparison against alternative families of nuclear luminosity histories.

More detailed dust-echo and reprocessing modeling can provide stronger geometric constraints when the driving optical/UV flare is well sampled. For example, convex-ring or torus-remnant models applied to AT~2019qiz \citep{Wu2025_AT2019qiz}, and multi-component disk-reprocessing and MIR dust-echo modeling of AT~2020nov \citep{Earl2025AT2020nov}, demonstrate how broader multiwavelength coverage can separate the primary emission, reprocessing structure, and dust response. In UGC~11487, the primary optical/UV driver was not observed, and the MIR coverage is limited to $W1$ and $W2$. The preferred two-component phenomenological $W1+W2$ description therefore constrains the presence of a structured MIR response, while the detailed dust geometry remains degenerate.

The radio measurements provide an important independent consistency check and were not incorporated into the MIR fitting procedure. The radio quantities used in the comparison include the cooling-break-corrected SED equipartition radii and the VLBA major-axis Gaussian scales reported by \citet{Golay2026}, whereas the MIR quantities are internal light-travel response scales inferred from the relative $W1$--$W2$ evolution. Figure~\ref{fig:mir_radio_scales} therefore compares characteristic sub-parsec scales defined by different observational diagnostics.

Recent JWST/MIRI spectroscopy of MIR-selected TDEs has revealed strong silicate emission together with compact high-ionization emission lines, showing that broadband MIR photometry alone cannot fully constrain the composition and geometry of the circumnuclear dust \citep{Masterson2025JWST}.

\section{Conclusions}
\label{sec:conclusions}

We investigated the decade-long MIR flare of the TDE candidate UGC~11487/WTP14adeqka using phase-resolved WISE/NEOWISE observations, empirical $W1$--$W2$ diagnostics, phenomenological decomposition,  and time-dependent radiative-transfer dust-echo modeling. We tested the temporal complexity of the observed response and its physical interpretation in terms of a STARS-driven nuclear flare illuminating an extended dust distribution. Our main conclusions are as follows:
\begin{enumerate}

\item UGC~11487/WTP14adeqka exhibits a decade-long MIR flare dominated
by dust-reprocessed emission. The flare shows a larger amplitude and
positive fluence in $W2$ than in $W1$, a delayed and broader $W2$
response, redder-when-brighter colour evolution, and flux--flux
hysteresis. These properties demonstrate that the MIR evolution is
strongly phase dependent.

\item The model-independent internal MIR timing diagnostics give a
preferred positive-fluence centroid delay of
$\Delta t_{\rm cent}=224\pm45$ d, corresponding to a characteristic
light-travel scale of $0.188\pm0.038$ pc. The shift-and-broaden analysis
\rev{gives a template-dependent shift scale of $c\tau=0.092\pm0.017$ pc. The fit tests the delayed-response scenario, but this shift is not a direct measurement of the physical delay.}

\item The effective colour-blackbody diagnostics show systematic
thermal evolution, with $T_{\rm eff}$ decreasing from $\sim900$ K
during the rise to $\sim500$--$600$ K during the decline. The effective
IR luminosity reaches $\sim3\times10^{43}$ erg s$^{-1}$, and the
integrated IR energy is of order
$(2.9$--$3.8)\times10^{51}$ erg.

\item The smooth joint $W1+W2$ modeling strongly
disfavors a single fixed-colour phenomenological component.
The preferred two-component phenomenological model gives
$\chi^2=21.06$ for $\nu=34$ and ${\rm BIC}=58.91$.
A third component follows the possible late-time $W2$
excess more closely, but is disfavored by
$\Delta{\rm BIC}_{3-2}=12.7$.
This comparison establishes the temporal complexity
of the MIR response without uniquely determining the
number of spatially distinct dust regions.

\item Two refined STARS-driven radiative-transfer solutions give $\chi^2=58.83$ (A) and 58.92 (B), each for nominal conditional $\nu=37$. Both use compact (0.25--0.50~pc) and extended (1.00--2.00~pc) graphite populations with fixed radial boundaries during refinement and satisfy the static-opacity criterion. They demonstrate the viability of a distributed dust response to a TDE-like flare, without uniquely identifying the source history or geometry.

\item The cooling-break-corrected radio SED radii are
$R_{\rm eq}=0.060$--$0.065$ pc, and the VLBA Gaussian major-axis
scales are $0.111$--$0.127$ pc \citep{Golay2026}. The compact MIR
shift scale, $c\tau=0.092\pm0.017$ pc, lies between these scales,
and the $W2$--$W1$ centroid scale,
$c\Delta t_{\rm cent}=0.188\pm0.038$ pc, traces a larger differential
MIR response scale. A possible late-time $W2$ excess occurs during the
radio-bright phase, with a formal significance of $2.06\sigma$ and a
one-sided bootstrap probability of $p=0.054$. The published radio-inferred energy is approximately \(1.7\)–\(2.2\%\) of the effective integrated IR energy. Powering the complete MIR flare through shocks would require a substantially larger kinetic-energy reservoir and efficient conversion into infrared emission. The available observations do not quantify this contribution or establish a physical association with the tentative late-time \(W2\) excess. \rev{The IR-based estimate gives $M_{\rm acc}\simeq0.032$--$0.21\,M_\odot$ and an illustrative disrupted-star mass scale of $M_{\star,\rm scale}\simeq0.065$--$2.1\,M_\odot$.}

\item The host galaxy is a moderately star-forming isolated barred disc
with $M_\star\simeq2\times10^{10}\,M_\odot$ and
${\rm SFR}\simeq3\,M_\odot\,{\rm yr}^{-1}$. The UV-to-radio SED
was modeled assuming an absent or weak persistent AGN component. The MIR variability requires a transient nuclear component superposed on the host.
\end{enumerate}

Overall, UGC~11487/WTP14adeqka demonstrates that phase-resolved WISE
$W1$--$W2$ diagnostics can recover characteristic sub-parsec response
scales and constrain the structure of dust reprocessing in obscured
accretion-powered nuclear transients with limited coverage of the
primary flare.

\begin{acknowledgements}
The research was supported by the National Research Foundation of Ukraine (project 2023.03/0188). This publication makes use of data products from  WISE and the NEOWISE mission, IRSA and NED which are funded by the NASA. We acknowledge the BHTOM platform \citep{Mikolajczyk2025BHTOM} for facilitating the exploratory inspection and visualization of the MIR variability of UGC~11487/WTP14adeqka. This work also uses observations obtained with the Samuel Oschin Telescope 48-inch and the 60-inch Telescope at the Palomar Observatory as part of the ZTF project. ZTF is supported by the National Science Foundation under Grants No. AST-1440341 and AST-2034437. The nuclear optical emission-line diagnostics cited in this work are taken from \citet{Golay2026} and are based on observations obtained with Binospec at the MMT Observatory, a joint facility of the Smithsonian Institution and the University of Arizona, under program SAO-13-24A (PI: Daichi Hiramatsu).
\end{acknowledgements}

\section{Data availability}
The WISE/NEOWISE single-exposure photometry used in this work is publicly available through the IRSA. The ZTF photometry is available through the public ZTF data services. The archival multiwavelength data used for the host-galaxy SED construction were obtained from public survey archives, including GALEX, Pan-STARRS1, 2MASS, WISE, IRAS, and NVSS. The radio constraints used for the MIR--radio comparison were taken from \citet{Golay2026}. The CIGALE code and the STARS fallback-rate library used in this work are publicly available \citep{Boquien2019, LawSmith2020STARS, LawSmith2020STARSZenodo}. The processed tables and analysis scripts are available from the corresponding author upon reasonable request. 
\begin{revisionblock}
The dust-echo implementation, documentation, archived Case~A illustration, and saved refined A/B analysis products are available in the \href{https://github.com/OlenaKompaniiets/tde-dust-echo/tree/b459f1f99ad38a42db59d798062b344c84343ab8}{version-pinned repository (commit b459f1f99ad38a42db59d798062b344c84343ab8)}. The \href{https://olenakompaniiets.github.io/tde-dust-echo/index.html}{interactive demonstration} presents the archived illustrative Case~A. The archive does not include every historical search product or all response caches needed to regenerate the full inference pipeline from scratch.
\end{revisionblock}

\bibliographystyle{aa} 
\bibliography{library} 
\makeatletter
\let\bibcite\@gobbletwo
\makeatother

\appendix
\nolinenumbers

\section{Smooth phenomenological component model}
\label{app:phenomenological_model_details}

The quantities defined in this appendix are effective diagnostics of the MIR
response and provide descriptive constraints on the WISE light-curve morphology.

The smooth $W1+W2$ phenomenological model used in Sect.~\ref{subsec:model_methods} was fitted to the baseline-subtracted WISE excess light curves. For each component $i$, the $W2$ excess flux was written as $\Delta F_{W2,i}(t)=A_i\,P_i(t)$, where $A_i$ is the $W2$ excess amplitude and $P_i(t)$ is a temporal profile
normalised to unity at its peak. The corresponding $W1$ excess flux was tied
to the same temporal profile through the temperature-dependent Planck ratio,
\begin{equation}
\Delta F_{W1,i}(t)=
A_i
\frac{B_\nu(\nu_{W1,{\rm rest}},T_i)}
{B_\nu(\nu_{W2,{\rm rest}},T_i)}
P_i(t).
\label{eq:app_w1_component}
\end{equation}

We adopted a smooth asymmetric Gaussian profile to provide a continuous description of the sparsely sampled WISE light curves,
\begin{equation}
P_i(t)=
\begin{cases}
\exp\left[-\frac{1}{2}
\left(\frac{t-t_{{\rm peak},i}}{\sigma_{{\rm rise},i}}\right)^2\right],
& t\leq t_{{\rm peak},i},\\[4pt]
\exp\left[-\frac{1}{2}
\left(\frac{t-t_{{\rm peak},i}}{\sigma_{{\rm decay},i}}\right)^2\right],
& t>t_{{\rm peak},i}.
\end{cases}
\label{eq:app_smooth_asymmetric_profile}
\end{equation}
The total model in each band was the sum of all components,
\begin{equation}
\Delta F_\nu(t)=
\sum_i \Delta F_{\nu,i}(t).
\label{eq:app_total_component_model}
\end{equation}
Because the quiescent MIR level had already been removed through the adopted baseline subtraction, the component model contains no additional constant background term.

\section{Time-dependent dust radiative-transfer model}
\label{app:hydro_clumpy_details}
\subsection{Fallback history and illuminating spectrum}

The selected STARS history is rescaled from $M_{\rm BH,ref}=10^6\,M_\odot$ to the adopted $M_{\rm BH}=10^{6.8}\,M_\odot$ at fixed stellar and encounter identifiers:
\begin{equation}
\begin{aligned}
s_M&=(M_{\rm BH}/M_{\rm BH,ref})^{1/2},\\
t&=s_Mt_{\rm ref},\\
\dot M_{\rm fb}(t)&=s_M^{-1}\dot M_{\rm fb,ref}(t/s_M).
\end{aligned}
\label{eq:rt_fallback_scaling}
\end{equation}
This preserves the integrated fallback mass. Source time zero is the unsmoothed fallback maximum; the tabulated rate is zero outside its supplied support. For $\tau_{\rm visc}>0$, causal smoothing \citep{Mockler2019} gives
\begin{equation}
\dot M_{\rm acc}(t)=\frac{1}{\tau_{\rm visc}}
\int_{-\infty}^{t}\dot M_{\rm fb}(t')
e^{-(t-t')/\tau_{\rm visc}}\,{\rm d}t'.
\label{eq:rt_accretion_convolution}
\end{equation}
The unit-integral kernel preserves mass over complete support; a finite-window tail is tracked separately. At $\tau_{\rm visc}=0$ the history is unsmoothed. This effective timescale does not separately resolve circularization and inward transport. An additional constant delay is fixed to zero because it is degenerate with $t_0$.

The isotropic source has $L_{\rm bol}=\eta\dot M_{\rm acc}c^2$ and $L_\lambda^{\rm src}=L_{\rm bol}s_\lambda$, with constant $\eta$ and the assumed, fixed $T_{\rm src}=3\times10^4$~K spectrum, normalized on the transport wavelength grid:
\begin{equation}
s_\lambda=\frac{B_\lambda(T_{\rm src})}
{\int B_{\lambda'}(T_{\rm src})\,{\rm d}\lambda'}.
\label{eq:rt_source_spectrum}
\end{equation}
No independent total-energy normalization or Eddington cap is imposed; mass loss and time-dependent efficiency are not separately modeled.

\subsection{Dust distribution and grain properties}

The reprocessor is an effective smooth, axisymmetric medium containing compact and extended radial populations, indexed by $j\in\{{\rm C,E}\}$. Their grain number densities, integrated over grain size, follow $n_j(r)\propto r^{-p_j}$ within $R_{{\rm in},j}\leq r\leq R_{{\rm out},j}$.
Both populations occupy the angular region $\mu|\leq\cos\theta_{\rm open},
\qquad \mu=\cos\theta,$ with no polar dependence of the prescribed density
inside this region. The polar opening angle $\theta_{\rm open}$ and observer inclination $i$ are measured from the symmetry axis. The central cavity, polar openings, and gap between the two radial populations contain no dust in the adopted configuration. Propagation is calculated along three-dimensional rays, while thermal quantities are represented on an
axisymmetric $(r,\mu)$ mesh.

Both populations have the fixed graphite size distribution $w(a)\propto a^{-q}$, normalized by $\int_{a_{\min}}^{a_{\max}}w(a)\,{\rm d}a=1$, with $q=3.5$, $a_{\min}=0.1\,\mu{\rm m}$, and $a_{\max}=1.0\,\mu{\rm m}$ \citep{Mathis1977}. Thus ${\rm d}n_j/{\rm d}a=n_j(r)w(a)$. We use the \texttt{Gra\_81.gz} absorption efficiencies for spherical graphite with the orientation-averaged $1/3$--$2/3$ prescription \citep{DraineLee1984,LaorDraine1993}. The same $C_{\rm abs}(\lambda,a)=\pi a^2Q_{\rm abs}(\lambda,a)$ enters heating, emission and attenuation; scattering is omitted.

The absolute dust normalization is defined by
\begin{equation}
\begin{aligned}
\overline C_\lambda&=\int_{a_{\min}}^{a_{\max}}w(a)C_{\rm abs}(\lambda,a)\,{\rm d}a,\\
\overline C_{\rm src}&=\int s_\lambda\overline C_\lambda\,{\rm d}\lambda,\\
\tau_j&=\overline C_{\rm src}\int_{R_{{\rm in},j}}^{R_{{\rm out},j}}n_j(r)\,{\rm d}r.
\end{aligned}
\label{eq:rt_tau_normalisation}
\end{equation}
The dimensionless $\tau_{\rm C}$ and $\tau_{\rm E}$ are source-spectrum-weighted radial absorption depths through the occupied angular region. They fix grain numbers and emitting cross-sections, and are distinct from $\tau_{\rm visc}$ and temporal lags. Finite-volume normalization and quadrature are documented in the version-pinned code.

\subsection{Radiative equilibrium and propagation times}

For each cell and grain radius $a$, radiative equilibrium \citep{Draine2003,Lu2016} requires
\begin{equation}
\begin{split}
\int C_{\rm abs}(\lambda,a)J_\lambda(\boldsymbol r,t_{\rm loc})\,{\rm d}\lambda
={}&\int C_{\rm abs}(\lambda,a)\\
&\times B_\lambda[T(\boldsymbol r,t_{\rm loc},a)]\,{\rm d}\lambda.
\end{split}
\label{eq:rt_radiative_equilibrium}
\end{equation}
Here $J_\lambda$ is mean intensity, including attenuated direct illumination and iterated secondary absorption and re-emission. The common $4\pi$ power factor cancels; single-grain emission is $P_\lambda=4\pi C_{\rm abs}B_\lambda(T)$. Equilibrium is assumed rapid: grain heat storage and stochastic fluctuations are omitted. The fixed photometric host adds no background thermal radiation field, so pre-flare values are not predictions of absolute quiescent dust temperatures.

For radius $r$ and observer direction $\hat{\boldsymbol n}$, pointing towards the observer, the source-retarded and arrival times are
\begin{equation}
\begin{aligned}
u&=t_{\rm loc}-r/c,\\
t_{\rm obs}&=t_0+(1+z)\left(t_{\rm loc}-\frac{\hat{\boldsymbol n}\cdot\boldsymbol r}{c}\right).
\end{aligned}
\label{eq:rt_observer_time}
\end{equation}
Secondary reprocessing adds causal path lengths, and escaping radiation is attenuated along the observer direction. Time differences are converted to days for comparison with MJD. The fitted $t_0$ aligns the unsmoothed fallback maximum, not an independently measured disruption date. A local thermal snapshot is simultaneous in $t_{\rm loc}$; an observed epoch combines emission from different local times.

\subsection{Static opacity and the sublimation check}

The grain population and absorption cross-sections are fixed. Temperatures are checked against the graphite sublimation threshold, approximately $1.79\times10^3$~K for the adopted hydrogen number density $n_{\rm H}=10^9\,{\rm cm}^{-3}$ \citep{BaskinLaor2018}; this is not the grain number density normalized by $\tau_j$. Above-threshold configurations are unsupported by the static-opacity solver, which does not evolve destruction, reformation or opacity changes. Unsupported status is neither an observational exclusion nor a physical zero-MIR solution.

\subsection{Synthetic photometry and the objective function}

The escaping spectrum is redshifted and converted to flux using the adopted luminosity distance, retaining its wavelength-dependent opacity and evolving temperatures.

The physical fit uses observed total flux densities at 22 NEOWISE epochs in both bands, with fixed host levels $F_{W1}^{\rm host}=8.072$~mJy and $F_{W2}^{\rm host}=5.215$~mJy from the first two NEOWISE epochs. These differ from the combined AllWISE+NEOWISE baseline used for empirical excess diagnostics and phenomenological fits. Host uncertainty is not propagated into the physical-model intervals. The canonical input uncertainties combine formal errors with an approximately 1.10~mJy additive floor; those stored effective uncertainties are used directly, without adding the floor again. The objective is
\begingroup
\boldmath
\begin{equation}
\chi^2_{\rm LC} =
\sum_{b\in\{W1,W2\}}\sum_i
\frac{
\left[
F_{b,i}^{\rm obs}
-F_b^{\rm host}
-F_b^{\rm RT}(t_i;\boldsymbol\theta,t_0)
\right]^2
}{
\sigma_{b,i,\rm eff}^{\,2}
}.
\label{eq:hydro_objective}
\end{equation}
\endgroup
For each physical configuration, $t_0$ is profiled over MJD~56500--57500. The likelihood treats the effective errors as independent. Empirical comparisons retain their stated baseline and time-window definitions.

\subsection{Parameter search}
\label{app:model_search_history}
Exploratory calculations guided the adopted graphite prescription but used different transfer approximations; they do not exclude silicate or mixed-grain models under the final solver. The retained A/B comparison uses one angular-transfer implementation, photometric dataset, error prescription, and baseline resolution.

The staged search used all 141 fallback files in the
adopted STARS set.
\rev{The discrete domain in Table~\ref{tab:rt_search_domain} contains 2025 combinations per history (285525 possible combinations), not that many completed calculations.}
Radial triples list
$(R_{\rm in}/{\rm pc},R_{\rm out}/{\rm pc},p)$.

\begin{table*}
\centering
\small
\caption{Discrete parameter domain of the staged
physical search. The design samples this domain through
the procedure described in the text.}
\label{tab:rt_search_domain}
\begin{tabular}{ll}
\hline
Parameter & Values \\
\hline
Compact radial population
& $(0.20,0.40,0)$, $(0.25,0.50,0)$,
$(0.30,0.60,0)$ \\
Extended radial population
& $(0.70,1.50,2)$, $(1.00,2.00,2)$,
$(1.50,3.00,2)$ \\
$\tau_{\rm C}$
& $0.2,\ 0.4,\ 0.6$ \\
$\tau_{\rm E}$
& $1.8,\ 2.6,\ 3.4$ \\
$\tau_{\rm visc}$ (d)
& $0,\ 50,\ 100,\ 200,\ 350$ \\
$\eta$
& $0.120,\ 0.135,\ 0.150,\ 0.165,\ 0.180$ \\
Opening angle and inclination
& $40^\circ$ each in the initial staged search \\
\hline
\end{tabular}
\end{table*}

A deterministic coarse design and local-neighbor exploration supplied the fine-resolution shortlist, including candidates with distinct fallback histories. Scheduling and duplicate-removal details are given in the version-pinned code.

Local refinement varies the six physical parameters listed in Sect.~\ref{subsec:model_methods}, with the fallback identifier and radial boundaries fixed.
The bounds are
$0.10\leq\eta\leq0.22$,
$100\leq\tau_{\rm visc}/{\rm d}\leq400$,
$0.2\leq\tau_{\rm C}\leq0.8$,
$1.2\leq\tau_{\rm E}\leq3.4$,
and $30^\circ$--$55^\circ$ for each angle.
The event epoch is profiled at each evaluation.

This finite, staged search identifies supported
candidate solutions within the specified domain.
Its coverage and ranking do not establish a global minimum
over continuous source and dust parameters. The retained branches are the independently refined \texttt{m3.0\_t0.0/0.850.dat} reference (A) and the staged-grid \texttt{m0.7\_t1.0/0.800.dat} candidate (B). Their shared radial boundaries were retained for local inference, not established as uniquely measured edges or an empty physical gap.

\subsection{Conditional parameter intervals}

For each selected fallback history and fixed radial
configuration, we constructed one-parameter profiles for
$\eta$, $\tau_{\rm visc}$, $\tau_{\rm C}$, $\tau_{\rm E}$,
the opening angle, and the inclination. The parameter of
interest was held fixed while the remaining physical
parameters and $t_0$ were reoptimised using an emulator
of the computed radiative-transfer responses.
Candidate endpoints near $\Delta\chi^2=1$ were evaluated
with the direct solver, with $t_0$ reoptimised for each
response. The other physical parameters in these direct
checks were fixed at their emulator-derived values.

Relative to the adopted direct-RT reference for
each case, the checked endpoints span
$\Delta\chi^2=0.96$--$1.04$ for Case~A and
$0.91$--$1.01$ for Case~B.
We report these bounds as approximate conditional
one-parameter intervals.
They describe parameter variation within the selected
source and dust prescription at the baseline numerical
resolution. The direct endpoint checks are not fully minimized direct-RT profiles and do not establish calibrated 68\% coverage or a complete uncertainty budget.

\subsection{Numerical and energy checks}

The reference fine calculation uses six radial cells
per dust population, six polar cells, \rev{41 grain-size nodes,} 144 wavelength samples,
and a 30-day transport time step.
The transport and observer integrations use 256 and 512
rays, respectively. \rev{The ray seed is 481.}
The wavelength grid spans $0.01$--$1000\,\mu{\rm m}$.
The source driver is sampled at 10-day resolution before
conservative integration into the transport time bins.
The observer response is followed to 15000 source-frame
days in the adopted time coordinate.

Numerical sensitivity was assessed at the adopted
Case~B physical parameters by independently refining the
angular sampling, halving the transport time step, doubling
the radial resolution, and doubling the wavelength sampling.
After reoptimising $t_0$, the corresponding changes relative
to the baseline $\chi^2$ were $-0.01$, $-0.06$, $+0.84$,
and $+0.01$, respectively.
The largest change arose from the radial refinement.
These comparisons quantify numerical sensitivity at the
reference configuration; the quoted parameter intervals
remain conditional on the baseline numerical resolution. Only $t_0$ was reoptimised; these tests do not measure fully refitted parameter shifts or interval coverage.

The energy accounting separates the source energy,
the directly absorbed source energy, the cumulative
thermal emission, secondary absorption, and eventual
escape. The cumulative thermal emission includes photons emitted
again after re-absorption and therefore contains internal
energy recycling.

For negligible stored thermal energy and a sufficiently
complete response interval, the escaping dust energy
satisfies
$E_{\rm IR,esc}=E_{\rm src,abs}$,
while the full budget satisfies
$E_{\rm src}=E_{\rm src,esc}+E_{\rm IR,esc}$.
Finite-window calculations additionally track radiation
whose absorption or escape remains pending.
The cumulative intrinsic dust emission is not substituted
for $E_{\rm IR,esc}$ in this comparison.

\section{Auxiliary optical checks}
\label{app:ztf}

\begin{figure}
\centering
\includegraphics[width=\columnwidth]{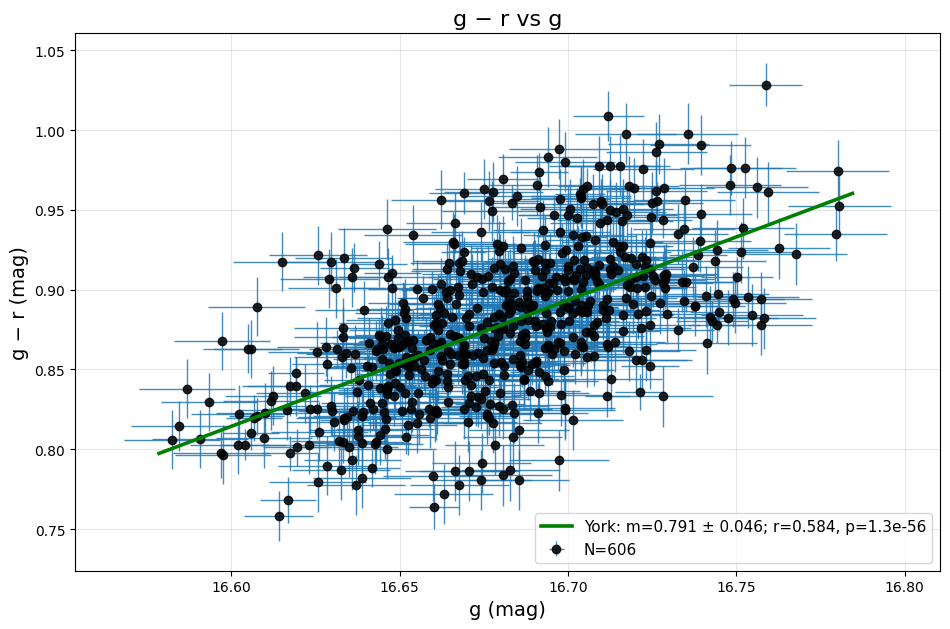}
\caption{
Post-flare ZTF $g-r$ colour--magnitude relation for UGC~11487 using the
$3\arcsec$ seeing cut. The ZTF data start after the main MIR rise and
maximum.
}
\label{fig:ztf_gr_colour}
\end{figure}

The available ZTF observations start after the main MIR rise and maximum of UGC~11487 and therefore cannot reconstruct the primary optical/UV flare or measure an optical--MIR lag. We use them only as a post-flare comparison dataset. Figure~\ref{fig:ztf_gr_colour} shows the $g-r$ colour--magnitude relation, while Table~\ref{tab:ztf_appendix} summarizes the colour and flux--flux checks.

The $g-r$ trend is significant, whereas the $r-i$ and flux--flux trends are weaker. Because the optical variability is small and likely host-dominated, these diagnostics are not used as physical constraints on the MIR flare.

\begin{table}
\centering
\caption{Post-flare ZTF colour and flux--flux diagnostics using a
$3\arcsec$ seeing cut.}
\label{tab:ztf_appendix}
\setlength{\tabcolsep}{3.2pt}
\begin{tabular}{l c c c}
\hline
Diagnostic & $N$ & Slope & $r$ \\
\hline
$g-r$ vs $g$ & 600 & $0.798\pm0.045$ & 0.585 \\
$r-i$ vs $r$ & 112 & $0.730\pm0.115$ & 0.552 \\
$\Delta F_g$ vs $\Delta F_r$ & 523 & $0.115\pm0.026$ & 0.249 \\
$\Delta F_r$ vs $\Delta F_i$ & 101 & $0.022\pm0.066$ & 0.135 \\
\hline
\end{tabular}
\end{table}

\section{Phenomenological response and component-model tests}
\label{app:two_component_model}

As an additional diagnostic, we fitted simple $W1$-to-$W2$ toy transfer functions in which the $W1$ excess was used as a proxy driver for the $W2$ excess. This test was used to assess the range of internal WISE response timescales. The Monte Carlo shift-and-broaden analysis used in the main text gives a compact effective shift of $\tau\simeq110\pm20$ d, corresponding to $c\tau\simeq0.092\pm0.017$ pc. The broader toy transfer-function tests give mean delays of $100$--160 d, equivalent to light-travel scales of $c\langle\tau\rangle\simeq0.08$--0.13 pc. The broad radial shell gives the lowest BIC, but several response functions reproduce the $W2$ morphology with comparable quality (Table~\ref{tab:toy_transfer_models}).

\begin{table}
\centering
\caption{Diagnostic toy $W1$-to-$W2$ transfer-function fits. The $W1$ excess
is used only as a proxy driver for the $W2$ excess.}
\label{tab:toy_transfer_models}
\setlength{\tabcolsep}{3.5pt}
\begin{tabular}{l c c c c}
\hline
Model & $k$ & $\Delta{\rm BIC}$ &
$\langle\tau\rangle$ & $c\langle\tau\rangle$ \\
 & & & d & pc \\
\hline
Broad radial shell   & 5 & 0.0 & 158 & 0.132 \\
Thin spherical shell & 3 & 3.4 & 101 & 0.084 \\
Gaussian response    & 4 & 4.7 & 104 & 0.088 \\
Inclined thin ring   & 4 & 5.7 & 105 & 0.088 \\
Top-hat response     & 4 & 6.4 & 101 & 0.084 \\
\hline
\end{tabular}
\tablefoot{
The models are diagnostic tests of response-timescale degeneracy and are used
as empirical response-shape comparisons.
}
\end{table}

\begin{figure}
\centering
\includegraphics[width=\columnwidth]{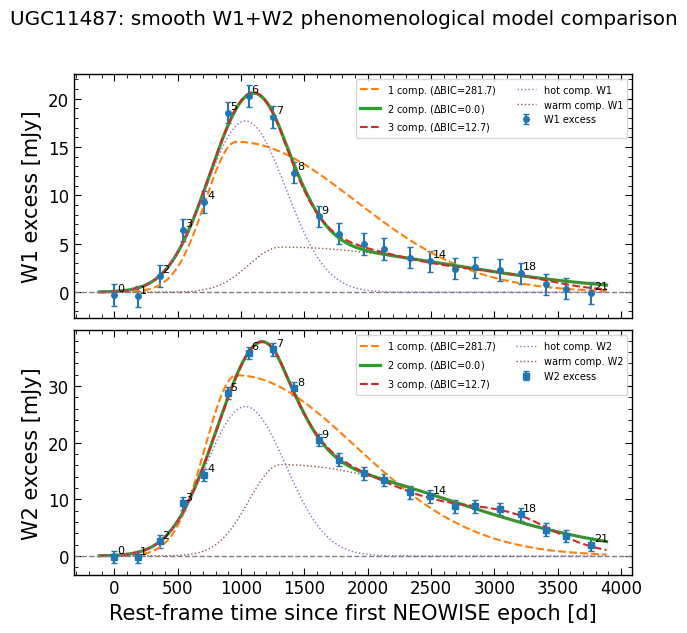}
\caption{
Smooth phenomenological $W1+W2$ component fits to the baseline-subtracted
WISE excess light curves. The three-component model follows the
possible late-time $W2$ excess more closely, but is disfavored by BIC for
the adopted effective error floor. It is included as a diagnostic of
possible late-time residual structure.} Dotted curves show the hot and warm
components of the preferred two-component phenomenological model.
\label{fig:smooth_component_model}
\end{figure}

Figure~\ref{fig:smooth_component_model} shows the smooth one-, two-, and three-component $W1+W2$ phenomenological fits to the baseline-subtracted WISE excess light curves. The one-component model does not reproduce the full $W1+W2$ morphology. The two-component model gives the preferred compact description for the adopted effective-error treatment. The three-component model is
included as a diagnostic of the possible late-time $W2$ excess.

Table~\ref{tab:two_component_parameters} gives the Monte Carlo parameter
estimates for the preferred smooth two-component phenomenological model. These
parameters describe an effective two-band MIR decomposition.

\begin{table}
\centering
\caption{Monte Carlo parameter estimates for the preferred smooth
two-component $W1+W2$ phenomenological model.}
\label{tab:two_component_parameters}
\renewcommand{\arraystretch}{1.18}
\setlength{\tabcolsep}{2.8pt}
\begin{tabular}{l@{\hspace{0.85em}}r@{\,}l@{\hspace{1.15em}}r@{\,}l}
\hline
Parameter &
\multicolumn{2}{c}{Hot component} &
\multicolumn{2}{c}{Warm component} \\
\hline
$A_{W2}$ & $26.5^{+2.4}_{-2.6}$ & mJy
& $16.1\pm1.0$ & mJy \\
$T$ & $878\pm46$ & K
& $524^{+31}_{-29}$ & K \\
$t_{\rm peak}$ & $1043^{+17}_{-19}$ & d
& $1317^{+86}_{-70}$ & d \\
$\sigma_{\rm rise}$ & $315^{+23}_{-19}$ & d
& $250^{+5}_{-0}$ & d \\
$\sigma_{\rm decay}$ & $309^{+26}_{-9}$ & d
& $1328^{+94}_{-78}$ & d \\
$B_{\nu}(W1)/B_{\nu}(W2)$
& $0.675^{+0.042}_{-0.044}$ &
& $0.289^{+0.036}_{-0.034}$ & \\
\hline
\end{tabular}
\tablefoot{
Values are the median parameters from Monte Carlo refitting, with
uncertainties given by the 16th--84th percentile range. $A_{W2}$ denotes
the peak $W2$ excess amplitude of each component. Times are rest-frame days
since the first NEOWISE epoch. The warm-component $\sigma_{\rm rise}$
reaches the imposed lower bound and is therefore a smoothing-scale
diagnostic with limited rise-time interpretation.
}
\end{table}

\section{Robustness and auxiliary checks}
\label{app:baseline_sensitivity}

Table~\ref{tab:baseline_sensitivity} shows how the model-independent
$W1$--$W2$ timing and fluence diagnostics depend on the adopted pre-flare
baseline. The numerical centroid scale changes with the baseline choice,
but $W2$ remains delayed and has a larger positive fluence than $W1$ in
all cases. The AllWISE 2010 points were used only to define and resample
the pre-flare baseline and were excluded from the flare fluence or energy
integrations.

The possible late-time $W2$ excess was assessed with a local
residual test based on the final 22-epoch WISE light curve. We fitted the
post-peak decline independently in $W1$ and $W2$ after excluding epochs
17--19. The fiducial model was an exponential-plus-constant profile,
\[
\Delta F(t)=C+A\exp[-(t-t_{\rm ref})/\tau_{\rm d}].
\]
The significance of the excluded epochs was calculated from their
weighted residuals, including the effective photometric uncertainties and
the empirical scatter of the fitted decline. We also performed 5000
parametric-bootstrap realizations of the smooth-decline null model, with
the decline refitted in every realization. Linear decline models and fit
ranges beginning at epochs 8, 9, and 10 were used to assess the sensitivity
of the result to the adopted decline prescription.

For the fiducial exponential-plus-constant model fitted from
epoch 9, the weighted residual over epochs 17--19 is
$1.49\pm0.72~{\rm mJy}$ in $W2$, corresponding to $2.06\sigma$.
The one-sided parametric-bootstrap probability is $p=0.054$. The
corresponding $W1$ residual is $0.59\pm0.65~{\rm mJy}$, or
$0.92\sigma$. For epochs 17--19, the $W2$ residual remains positive
for all tested fit ranges and decline models, with significances of
approximately $1.4$--$2.1\sigma$. The largest positive $W2$ residuals
occur at epochs 17 and 18, while epoch 19 is consistent with the fitted
decline. These results indicate a possible late-time $W2$ excess with
modest statistical significance.

\begin{figure}
\centering
\includegraphics[width=\columnwidth]
{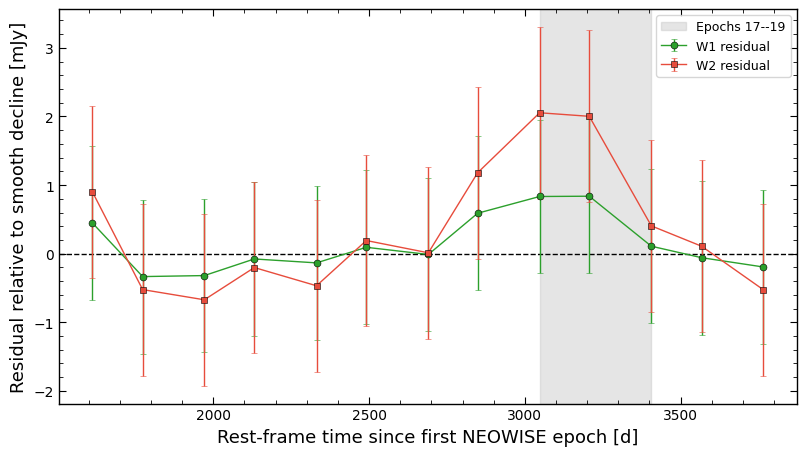}
\caption{Residuals of the post-peak $W1$ and $W2$ excess light
curves relative to the fiducial exponential-plus-constant decline fitted
after excluding epochs 17--19. The shaded region marks the interval tested
for a possible late-time excess. The $W2$ residual is positive, particularly at epochs 17 and 18. The weighted residual over the three epochs has a formal significance of $2.06\sigma$, with a one-sided parametric-bootstrap probability of $p=0.054$.}
\label{fig:late_W2_residuals}
\end{figure}

\begin{table}
\centering
\small
\caption{Baseline sensitivity of the $W1$--$W2$ timing and fluence
diagnostics. Centroid offsets are rest-frame values.}
\label{tab:baseline_sensitivity}
\setlength{\tabcolsep}{2.8pt}
\renewcommand{\arraystretch}{1.10}
\begin{tabular}{@{}l@{\hspace{4pt}}c@{\hspace{4pt}}c@{\hspace{4pt}}c@{\hspace{4pt}}c@{\hspace{4pt}}c@{}}
\hline
Baseline &
$F_{W1,0}$ &
$F_{W2,0}$ &
$\Delta t_{\rm cent}$ &
$c\Delta t_{\rm cent}$ &
$\Phi_{W2}/\Phi_{W1}$ \\
& mJy & mJy & d & pc & \\
\hline
NEOWISE 0--1
& 8.072 & 5.215 & 206.6 & 0.173 & 2.16 \\
AllWISE+NEOWISE
& 8.453 & 5.428 & 233.8 & 0.196 & 2.25 \\
AllWISE 2010 only
& 8.786 & 5.672 & 256.6 & 0.215 & 2.33 \\
\hline
\end{tabular}
\tablefoot{
The fiducial main-text value,
$\Delta t_{\rm cent}=224\pm45$ d and
$c\Delta t_{\rm cent}=0.188\pm0.038$ pc, comes from the Monte
Carlo calculation in which the combined baseline is bootstrapped. The fixed
combined-baseline value listed here is an auxiliary baseline-sensitivity
quantity.
}
\end{table}

Table~\ref{tab:error_floor_sensitivity} shows the dependence of the smooth component comparison on the adopted effective error floor. The one-component model remains strongly disfavoured for all tested floors. The three-component model is preferred only for the lowest tested value, $\sigma_{\rm floor}=0.50~{\rm mJy}$, where small residuals, including the possible late $W2$ shoulder, receive greater statistical weight.

\begin{table}
\centering
\small
\caption{Sensitivity of the smooth $W1+W2$ component comparison to the
effective error floor. $\Delta{\rm BIC}_{1-2}$ and
$\Delta{\rm BIC}_{3-2}$ are relative to the two-component model.}
\label{tab:error_floor_sensitivity}
\setlength{\tabcolsep}{1.6pt}
\renewcommand{\arraystretch}{1.08}
\begin{tabular}{@{}c c c c c c c@{}}
\hline
$\sigma_{\rm floor}$ &
$\chi^2_{\nu,1}$ &
$\chi^2_{\nu,2}$ &
$\chi^2_{\nu,3}$ &
$\Delta{\rm BIC}_{1-2}$ &
$\Delta{\rm BIC}_{3-2}$ &
BIC pref. \\
mJy & & & & & & \\
\hline
0.50 & 39.40 & 2.97 & 2.41 & 1416.6 & -12.1 & 3 comp. \\
0.80 & 15.55 & 1.17 & 0.95 & 548.0 & 6.7 & 2 comp. \\
1.10 & 8.25 & 0.62 & 0.50 & 281.7 & 12.7 & 2 comp. \\
1.50 & 4.45 & 0.33 & 0.27 & 143.1 & 15.4 & 2 comp. \\
2.00 & 2.50 & 0.19 & 0.15 & 72.3 & 17.0 & 2 comp. \\
\hline
\end{tabular}
\tablefoot{
The adopted analysis uses $\sigma_{\rm floor}=1.10$ mJy. The
one-component model is always strongly disfavored; the two- versus
three-component distinction is sensitive only at the smallest tested floor.
}
\end{table}

We also tested whether the model comparison depends on the
functional form of the adopted uncertainty by repeating the smooth
$W1+W2$ one-, two-, and three-component fits with the 10\% flux-dependent
prescription
\[
\sigma_{\rm eff,10\%}
=
\left[
\sigma_{F_\nu}^{2}
+
(0.10F_\nu)^{2}
\right]^{1/2},
\]
where $F_\nu$ is the observed total epoch flux density. The model
ranking is unchanged: the one-component model remains strongly
disfavored, while the two-component model remains preferred over the
three-component model by BIC.

\begin{table}
\centering
\footnotesize
\caption{Sensitivity of the smooth $W1+W2$ model comparison to
the uncertainty prescription.}
\label{tab:error_model_sensitivity}
\setlength{\tabcolsep}{4.2pt}
\renewcommand{\arraystretch}{1.05}
\begin{tabular}{@{}lcccccc@{}}
\hline
Error model &
$\chi^2_{\nu,1}$ &
$\chi^2_{\nu,2}$ &
$\chi^2_{\nu,3}$ &
$\Delta{\rm BIC}_{1-2}$ &
$\Delta{\rm BIC}_{3-2}$ &
Pref. \\
\hline
Fixed $1.10$ mJy
& 8.25 & 0.62 & 0.50 & 281.7 & 12.7 & 2 comp. \\
10\% total flux
& 2.50 & 0.47 & 0.34 & 62.7 & 13.0 & 2 comp. \\
\hline
\end{tabular}
\tablefoot{
The fixed floor is the fiducial prescription. The 10\%
total-flux case follows the flux-dependent treatment motivated by
\citet{Masterson2024}. Positive $\Delta{\rm BIC}$ values favor the
two-component model.}
\end{table}

Table~\ref{tab:radio_travel_time} lists the illustrative
travel-time velocities used for the MIR--radio comparison discussed in
Sect.~\ref{subsec:radio_results}. The radio epochs are taken from the
published observations of WTP14adeqka by \citet{Golay2026}. The midpoint
of the possible late-time $W2$ excess is defined as the median MJD of WISE
epochs 17--19. These estimates compare the characteristic MIR
dust-response scales with the independently measured radio-emitting
region.

\begin{table*}
\centering
\caption{Illustrative MIR--radio travel-time comparison.}
\label{tab:radio_travel_time}
\setlength{\tabcolsep}{5pt}
\renewcommand{\arraystretch}{1.18}

\begin{tabular*}{\textwidth}{@{\extracolsep{\fill}}lccc}
\hline
Reference event &
Compact shift &
$W2$--$W1$ centroid &
Literature shell \\
&
$R=0.092$ pc &
$R=0.188$ pc &
$R=0.33$ pc \\
\hline

First radio detection &
{ $\beta_{\rm req}\sim0.077$} &
{ $\sim0.16$} &
{ $\sim0.27$} \\

Radio bright epoch &
{ $\beta_{\rm req}\sim0.046$} &
{ $\sim0.094$} &
{ $\sim0.17$} \\

$W2$ late-excess interval midpoint &
{ $\beta_{\rm req}\sim0.040$} &
{ $\sim0.082$} &
{ $\sim0.14$} \\

VLBA epoch 1 &
{ $\beta_{\rm req}\sim0.035$} &
{ $\sim0.072$} &
{ $\sim0.13$} \\

VLBA epoch 2 &
{ $\beta_{\rm req}\sim0.031$} &
{ $\sim0.064$} &
{ $\sim0.11$} \\

\hline
\end{tabular*}

\tablefoot{
The listed values give the characteristic propagation velocity
required to reach each MIR scale from the adopted reference epoch,
2015 April 14 (${\rm MJD}=57126$), by the corresponding radio or
late-MIR epoch. The radio epochs are taken from \citet{Golay2026}.
The midpoint of the possible late-time $W2$ excess is defined as the
median MJD of WISE epochs 17--19,
${\rm MJD}=59911.00425$. The observed-frame elapsed times are converted
to the source rest frame using $z=0.018950$ before calculating
$\beta_{\rm req}=R/(c\,\Delta t_{\rm rest})$. The compact-shift scale is
derived from the shift-and-broaden analysis, and the $W2$--$W1$ centroid
scale from the positive-fluence centroid offset. The literature-shell
scale is the published dust-shell radius,
$R_{\rm dust,shell}=0.33^{+0.13}_{-0.05}$ pc
\citep{Masterson2024}. The velocity associated with the late-$W2$
interval is an illustrative travel-time estimate and does not imply an
independently established launch time or a causal connection with the
radio outflow.}
\end{table*}

\section{Aperture-matched host-galaxy photometry and SED modeling}
\label{app:host_sed_methods}

To characterize the time-averaged host-galaxy emission of UGC~11487, we performed aperture-matched photometry using archival GALEX, Pan-STARRS1, 2MASS, WISE, IRAS, and NVSS data. The methodology for aperture photometry described in \cite{Kompaniiets2026}. The purpose of this analysis is to provide the host-galaxy context for the MIR flare. The resulting UV--radio SED was used to estimate the stellar mass, star formation rate, dust attenuation, and dust luminosity under the adopted absent or weak persistent AGN assumption.

For the UV, optical, near-infrared, and WISE $W1$--$W2$ bands, we adopted a common full-galaxy elliptical aperture centered on UGC~11487, with semi-major axis $a=45\arcsec$, semi-minor axis $b=30\arcsec$, and position angle $PA=40^\circ$. This aperture encloses the main optical body of the galaxy while limiting the contribution from the surrounding background. For the lower-resolution WISE W3 and W4 images, we used circular apertures with radii of $50\arcsec$ and $55\arcsec$, respectively. The NVSS 1.4 GHz flux density was measured with a circular aperture of radius $65\arcsec$, appropriate for the NVSS beam size.

A local background was estimated and subtracted in each band before aperture summation. A reference contaminant mask was constructed from the Pan-STARRS1 $r$-band image and reprojected to the Pan-STARRS1 $g,r,i$, 2MASS $J,H,K_s$, and WISE W1--W3 images. The mask was dilated according to the band-dependent point-spread function. Masked pixels inside the aperture were replaced by a smooth local estimate before the final aperture summation. No contaminant mask was applied to GALEX FUV/NUV, WISE W4, or NVSS because of the lower angular resolution and the different morphology of the diffuse emission.

The aperture sums were converted to flux densities using the appropriate survey calibrations: AB calibrations for GALEX and Pan-STARRS1, Vega zero-points for 2MASS and WISE where required, surface-brightness conversion for images in MJy sr$^{-1}$, and beam-aware conversion for NVSS maps in Jy beam$^{-1}$. The UV, optical, and near-infrared fluxes were corrected for Galactic foreground extinction using $E(B-V)=0.0643$ and a standard $R_V=3.1$ extinction curve. All final broadband flux densities and uncertainties were converted to mJy before being used in the SED modeling.

The Pan-STARRS1 $y$-band measurement was excluded from the final SED fit. An aperture-growth diagnostic showed an anomalously high $y/i$ flux ratio, $F_y/F_i\simeq4.9$, at the adopted aperture, inconsistent with the smoother behaviour of the $r/g$ and $i/r$ ratios. We therefore treated the $y$-band flux as affected by calibration, background, or large-scale image-structure systematics.

IRAS photometry was treated separately because of the poorer angular resolution and the strong cirrus/confusion background at the position of UGC~11487. For the 12, 25, 60, and 100 $\mu{\rm m}$ maps, we used beam-aware circular-aperture photometry in MJy sr$^{-1}$ images with local annular background subtraction. The uncertainty was estimated empirically from random apertures in the surrounding field. The 12, 25, and 60 $\mu{\rm m}$ measurements had low or negative random-aperture significance and were therefore used as conservative $3\sigma$ upper limits. \rev{The 100 $\mu{\rm m}$ measurement is also treated as an upper limit; its nominal significance is ${\rm S/N}\simeq3.3$.}

The resulting broadband SED was modelled with CIGALE. The fit used the aperture-matched UV--radio photometry to describe the time-averaged host-galaxy emission. The WISE W1 and W2 points used in this SED fit represent the integrated host-galaxy flux level and were excluded from the time-dependent MIR flare modeling. \rev{The SED modeling characterizes the persistent stellar, star-forming, and dust components under this assumption.} The best-fitting SED and the derived host-galaxy parameters are presented in Sect.~\ref{subsec:sed_results}.

\begin{figure}[tp]
\centering

\begin{minipage}[t]{0.48\columnwidth}
  \centering
  \includegraphics[width=\linewidth]{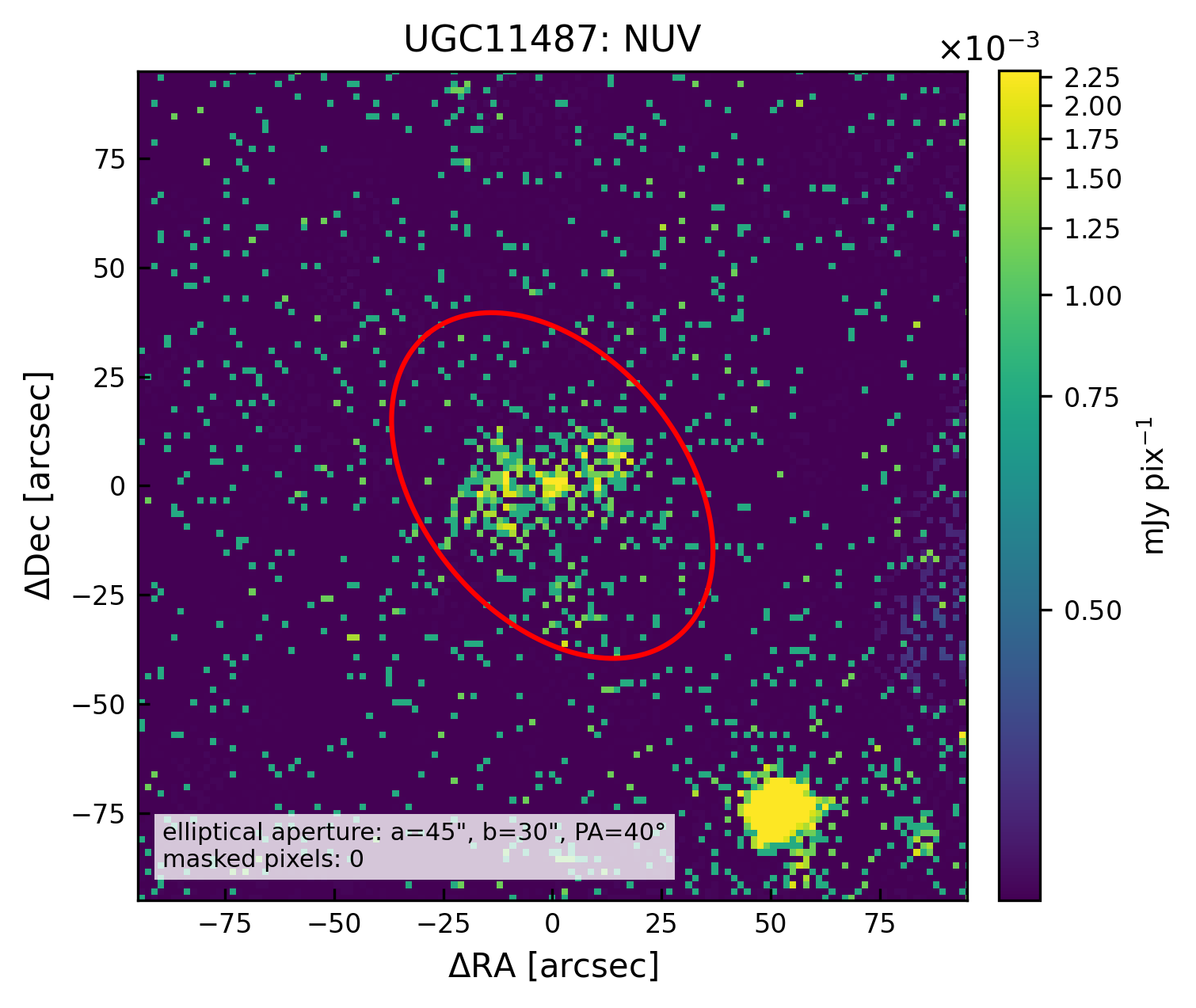}\par
  {\small (a) GALEX NUV\par}
\end{minipage}\hfill%
\begin{minipage}[t]{0.48\columnwidth}
  \centering
  \includegraphics[width=\linewidth]{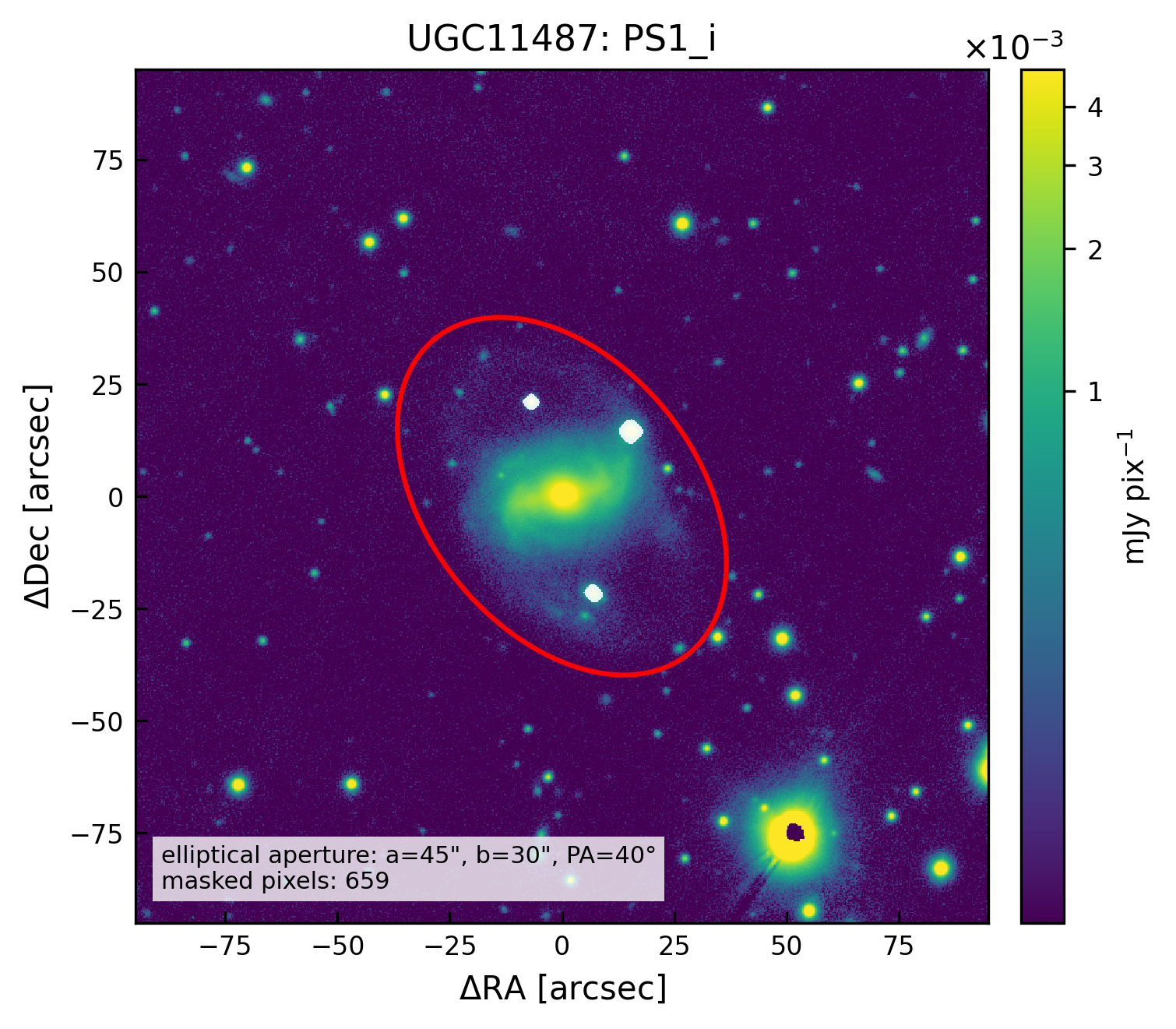}\par
  {\small (b) Pan-STARRS1 $i$-band\par}
\end{minipage}

\par\medskip

\begin{minipage}[t]{0.48\columnwidth}
  \centering
  \includegraphics[width=\linewidth]{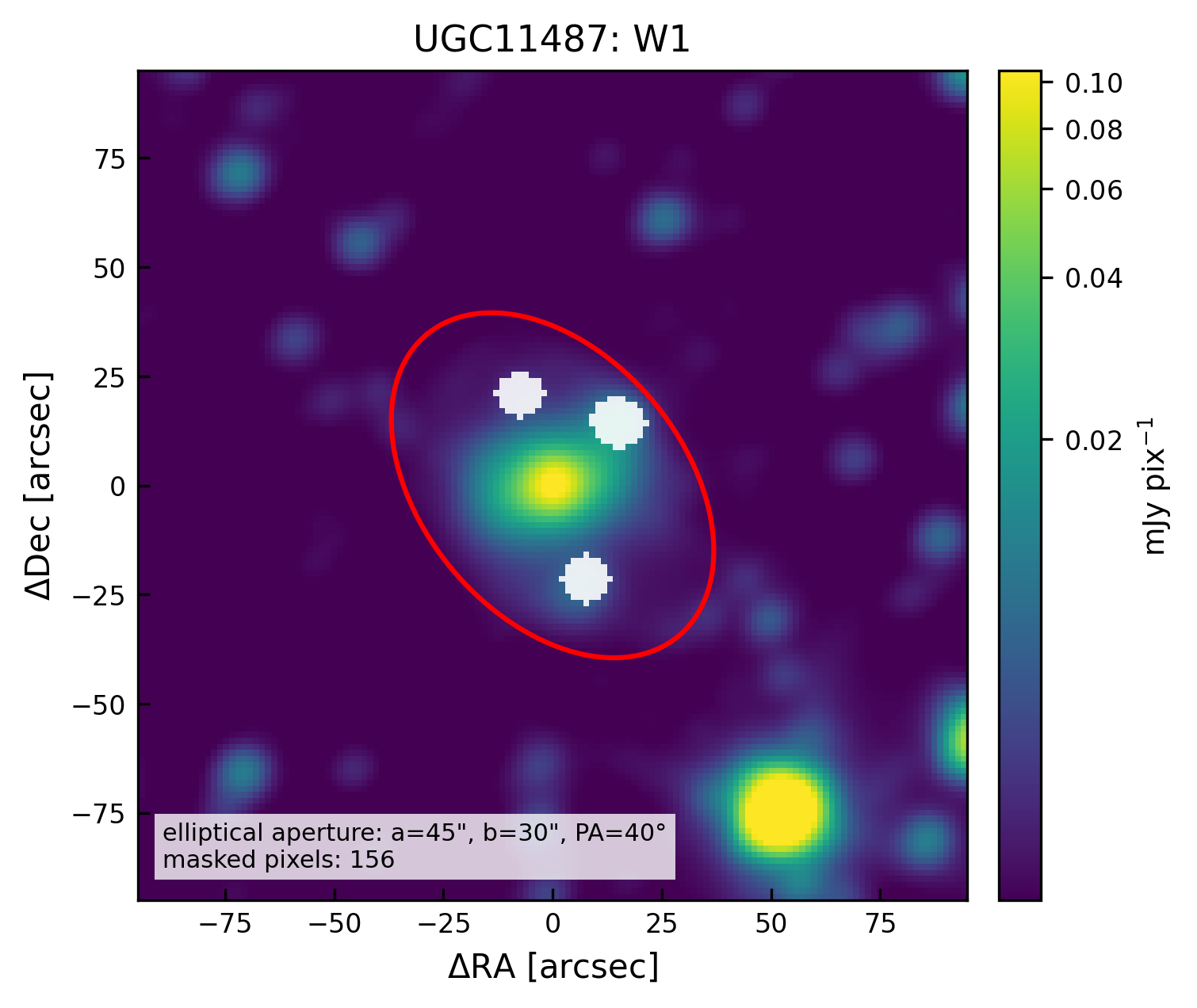}\par
  {\small (c) WISE W1\par}
\end{minipage}\hfill%
\begin{minipage}[t]{0.48\columnwidth}
  \centering
  \includegraphics[width=\linewidth]{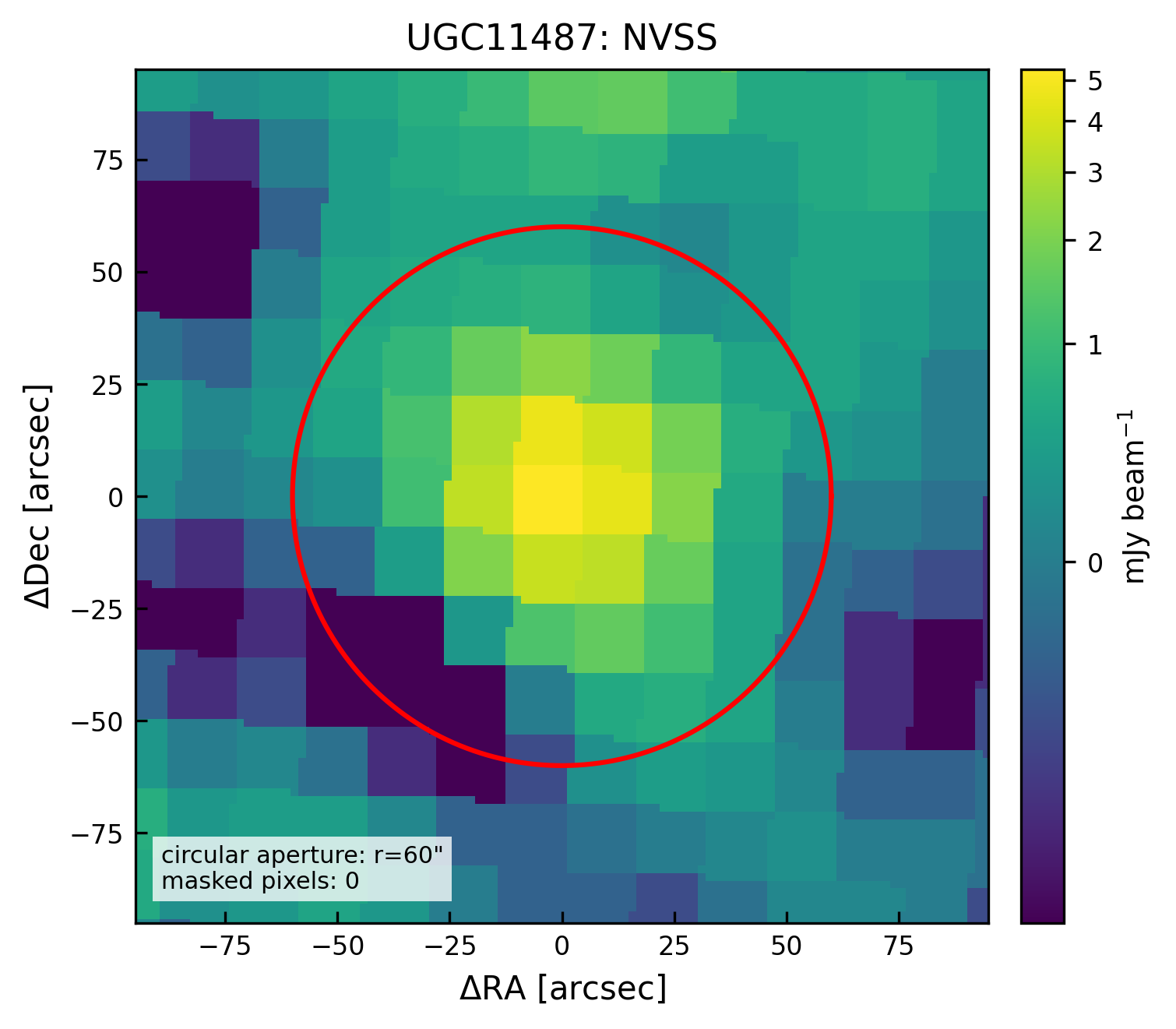}\par
  {\small (d) NVSS 1.4 GHz\par}
\end{minipage}

\caption{
Examples of multiwavelength images used to validate the aperture-matched
host-galaxy photometry of UGC~11487: (a) GALEX NUV, (b) Pan-STARRS1
$i$-band, (c) WISE W1, and (d) NVSS 1.4 GHz. All panels show a
$190\arcsec\times190\arcsec$ field centered on the galaxy. The red contour
marks the aperture used for the corresponding integrated flux measurement:
the common full-galaxy elliptical aperture for the UV, optical, and WISE W1
images, and the circular low-resolution aperture for the NVSS map. White
regions inside the aperture indicate masked contaminating sources.
}
\label{fig:ugc11487_multiwavelength_photometry}
\end{figure}

\section{CIGALE input parameters for SED modeling}
\label{app:CIGALE_input}
We modeled the broadband spectral energy distribution (SED) of UGC~11487 with CIGALE using the \texttt{pdf\_analysis} method. The input photometry covers the UV-to-radio wavelength range. \rev{The four IRAS bands (12, 25, 60, and 100 $\mu{\rm m}$) are specified as upper limits for the CIGALE likelihood calculation.}

The SED was constructed using the following sequence of \textsc{CIGALE} modules \citep{bruzual2003}: \texttt{sfhdelayed}, \texttt{bc03}, \texttt{nebular}, \texttt{dustatt\_modified\_starburst}, \texttt{dale2014}, \texttt{radio}, and \texttt{redshifting}. The stellar emission was computed using the \cite{bruzual2003} stellar population synthesis models with a \cite{chabrier2003} initial mass function. Nebular emission was included following \citet{inoue2011}. Dust attenuation was modeled with a modified starburst attenuation law based on \citet{calzetti2000}, and the infrared dust emission was described using the templates of \citet{dale2014}. The radio continuum was modeled through the standard \textsc{CIGALE} radio prescription, which relies on the FIR/radio correlation \citep[e.g.][]{helou1985, yun2001}. The \texttt{redshifting} module accounts for the source redshift and includes intergalactic medium absorption following \citet{meiksin2006}. The adopted parameter grid is given in Table~\ref{tab:cigale_grid_ugc11487}.

\begin{table}[tp]
\small
\centering
\caption{CIGALE modules and parameter grid used for the SED modeling of UGC~11487.}
\label{tab:cigale_grid_ugc11487}
\begin{tabular}{l@{\hspace{2em}}l}
\hline
Parameter & Adopted values \\
\hline
\multicolumn{2}{c}{\texttt{sfhdelayed}} \\
\hline
$\tau_{\rm main}$ [Myr] & 500, 1000, 2000, 4000, 6000 \\
$age_{\rm main}$ [Myr] & 8000, 9000, 10000, 11000, 12000 \\
$\tau_{\rm burst}$ [Myr] & 10 \\
$age_{\rm burst}$ [Myr] & 10 \\
$f_{\rm burst}$ & 0.0 \\
normalize & True \\
\hline
\multicolumn{2}{c}{\texttt{bc03}} \\
\hline
IMF & Chabrier \\
metallicity & 0.02 \\
separation age [Myr] & 10 \\
\hline
\multicolumn{2}{c}{\texttt{nebular}} \\
\hline
$\log U$ & $-2.0$, $-3.0$, $-3.5$, $-4.0$ \\
$Z_{\rm gas}$ & 0.02 \\
$n_{\rm e}$ [$\mathrm{cm^{-3}}$] & 100 \\
$f_{\rm esc}$ & 0.0 \\
$f_{\rm dust}$ & 0.0 \\
line width [$\mathrm{km\,s^{-1}}$] & 300 \\
\hline
\multicolumn{2}{c}{\texttt{dustatt\_modified\_starburst}} \\
\hline
$E(B-V)_{\rm lines}$ & 0.3, 0.4, 0.5, 0.6, 0.7, 0.8, 1.0 \\
$E(B-V)$ factor & 0.44 \\
UV bump wavelength [nm] & 217.5 \\
UV bump width [nm] & 35.0 \\
UV bump amplitude & 0.0 \\
power-law slope $\delta$ & 0.2, 0.1, 0.0, $-0.1$, $-0.25$, $-0.5$ \\
emission-line extinction law & MW \\
$R_V$ & 3.1 \\
\hline
\multicolumn{2}{c}{\texttt{dale2014}} \\
\hline
$f_{\rm AGN}$ & 0.0, 0.001, 0.01, 0.05, 0.1 \\
$\alpha$ & 1.5, 1.625, 1.75, 1.875, 2.0, 2.25 \\
\hline
\multicolumn{2}{c}{\texttt{radio}} \\
\hline
$q_{\rm IR,SF}$ & 2.25, 2.58, 2.75 \\
$\alpha_{\rm SF}$ & 0.8 \\
$R_{\rm AGN}$ & 0 \\
$\alpha_{\rm AGN}$ & 0 \\
\hline
\multicolumn{2}{c}{\texttt{redshifting}} \\
\hline
redshift & from the input file \\
\hline
\end{tabular}
\end{table}

\end{document}